\documentclass[preprint]{aastex}
\usepackage{emulateapj5}

\newcommand{\hh}{$^{\mathrm h}$}
\newcommand{\mm}{$^{\mathrm m}$}

\received{March 2003}
\accepted{April 2003}
\slugcomment{Submitted to  {\em The Astronomical Journal}.}
\shortauthors{Tinney, Burgasser \& Kirkpatrick}
\shorttitle{Infrared T dwarf Parallaxes}
\begin{document}

\title{Infrared Parallaxes for Methane T dwarfs\altaffilmark{1}}
\author{C.G. Tinney\altaffilmark{2}, 
        Adam  J. Burgasser\altaffilmark{3,4}, 
        J. Davy. Kirkpatrick\altaffilmark{5}}
\altaffiltext{1}{Based on observations obtained at the 
European Southern Observatory, Chile. Programmes
65.L-0061, 66.C-0404, 67.C-0029, 68.C-0004 \& 69.C-0044.}
\altaffiltext{2}{Anglo-Australian Observatory, PO Box 296. 
                Epping. 1710 Australia. {\tt cgt@aaoepp.aao.gov.au}}
\altaffiltext{3}{UC Los Angeles, 8371 Mathematical Sciences, CA. 90095. USA. {\tt adam@astro.ucla.edu}}
\altaffiltext{4}{Hubble Fellow}
\altaffiltext{5}{Infrared Processing \& Analysis Center, Caltech, Pasadena, CA. 91125  USA. {\tt davy@ipac.caltech.edu}}

\label{firstpage}

\begin{abstract}
We report final results from our 2.5 year infrared parallax
program carried out with the European Southern Observatory
3.5m New Technology Telescope and the SOFI infrared camera.
Our program targeted precision astrometric observations of
ten T type brown dwarfs in the J band. Full astrometric solutions (including
trigonometric parallaxes) for nine T dwarfs are
provided along with proper motion solutions for a further object. 
We find that HgCdTe-based infrared cameras
are capable of delivering precision differential astrometry.
For T dwarfs, infrared observations are to be greatly
preferred over the optical, both because
they are so much brighter in the infrared, and because their prominent
methane absorptions lead to similar effective wavelengths
through the J-filter for both target and reference stars, which in turn
results in a dramatic reduction in differential colour refraction effects.
We  describe a technique for robust bias estimation and linearity
correction with the SOFI camera, along with an upper limit
to the astrometric distortion of the SOFI optical train.
Colour-magnitude and spectral-type-magnitude diagrams for
both L  and T dwarfs are presented which show complex and
significant structure, with major import for luminosity function 
and mass function work on T dwarfs. Based on the width of the
early L dwarf and late T dwarf colour magnitude diagrams,
we conclude the brightening of early T dwarfs in the J passband
(the ``early T hump'') is not an age effect, but due to 
the complexity of brown dwarf cooling curves. Finally, empirical estimates
of the ``turn on'' magnitudes for methane absorption in
field T dwarfs and in young stars clusters are provided.
These make the interpretation of the T6 dwarf $\sigma$\,Ori\,J053810.1-023626
as a $\sigma$\,Ori member problematic.
\end{abstract}

 
\section{Introduction -- Methane T-type Brown Dwarfs}

Numerous examples of the field counterparts to the extremely cool methane brown dwarf
Gl\,229B \citep{nak1995} are now known \citep{st1999,bu1999,bu2000a,bu2000b,le2000,ts2000,cu2000}. 
These objects are now uniformly classified as ``T dwarfs'' \citep{bu2002a,ge2002},
and  have such low photospheric temperatures (800-1300K), that 
their photospheres are dominated by the effects of dust 
and methane formation \citep{all2001}, neither of which are amenable to simple
modeling. The discovery of sizable numbers
of T dwarfs, means that we are now in a position to use direct
trigonometric parallax observations to {\em empirically}
determine the loci of T dwarf cooling curves, rather than relying on models. 
The discovery of several T dwarfs by SDSS with spectra bridging
the L  and T  spectral types (e.g. \citealt{le2002,ge2002})  means we
are also in a position to empirically determine where on these brown dwarf 
cooling curves  the L-T transition occurs.

Trigonometric parallaxes are also essential to understanding the
space density of T dwarfs. Luminosity function estimates
for T dwarfs (eg. \citealt{burg_thesis}) are currently based 
on limited parallaxes  and assumptions about object binarity.
(Recent programs targeting more L  and T dwarfs \citep{mbb1999,ko1999,reid2001,bu2003a,close2003}, 
indicate that $\sim$10-20\% of objects observed 
in sufficient detail are found to be binary.) 
Luminosity functions based on currently available colour-magnitude
relations will therefore be problematic at best. Trigonometric
parallaxes are therefore required to determine the {\em actual} luminosities
of these objects and indicate whether they are single or binary, so that more
meaningful luminosity functions for T dwarfs can be constructed.

\section{Parallaxes and the Infrared}

Traditional parallax techniques based on photography are completely
unable to target objects as faint and red as T dwarfs. CCD parallax work 
in the optical at the USNO, ESO and Palomar  \citep{mo1992,da2002,t96,t95,t93} 
have shown that parallaxes can be obtained for
objects as faint as I=18-19 at distances $\la$70\,pc.
However, this still leaves the T dwarf class of objects (with I$\ga$21)
unobservable. To date only a few of the very brightest and closest
T dwarfs have proved tractable for CCD parallax work \citep{da2002}.

Over the last two years, therefore, we have been extending optical CCD
astrometric techniques into the infrared, where 
the J$<$16 magnitudes of most of the detected T dwarfs make 
significant progress possible.
Indeed, there are several reasons to {\em prefer} the infrared
for high precision astrometry. 
First, the effects of differential colour refraction
(the different amount of refraction the atmosphere produces
in red target stars, compared to blue reference stars, see \citealt{mo1992}) are reduced
by working at longer wavelengths. Second, because T dwarfs suffer
methane absorption at the red end of their J- and H-band spectra,
their effective wavelengths through a J- or H-filter are much closer
to that of a typical background reference star, than is the case
in the optical. These effects combined mean
that the stringent requirements on maintaining
control of observations at constant hour angles (at least for T dwarfs)
is not present in the infrared (cf. Section \ref{sec_dcr}). This {\em considerably} increases the
flexibility and efficiency of infrared parallax observing, over the optical.
Third, seeing improves in the infrared, leading to smaller images,
smaller amounts of differential seeing, and so higher astrometric precision.
And finally, T dwarfs show {\em much} greater contrast to
sky in the near-infrared than in the optical. 

Infrared parallax observations have been pioneered by \citet{jo2000}, who
targeted the extremely active (and unfortunately at 76\,pc also quite
distant) late M dwarf PC0025+0447, as well as the nearby M dwarf VB10.
The USNO also has an infrared astrometric program in operation,
from which published results are expected shortly \citep{vr2002}.

\section{Observations \& Sample}

Observations were carried out at 7 epochs over the period
2000 April 17 to 2002 May 30. At each epoch, observations were
carried out on either two half- or two full-nights. All observations
were obtained with the SOFI infrared camera on the European Southern 
Observatory (ESO) 3.5m New Technology Telescope (NTT).
SOFI was used in its ``large field'' mode in which it provides a 4.92\arcmin$\times$4.92\arcmin\
field-of-view with 0.28826\arcsec\ pixels (cf. Section \ref{scale}). 

Exposures of each target were acquired with a fixed dither pattern (Fig. \ref{jitter}) as eight
120s exposures though the SOFI J filter. The exposure pattern was designed so that
this 16 minutes of dithered exposure time sampled many different inter-pixel spacings.
As much as was feasible (given observing time constraints)
we attempted to acquire all epoch observations at the same hour angle as
the very first epoch observation acquired, so as to minimise 
DCR effects. Each epoch observation was also carried out with a specified
reference star positioned within a few pixels of its position when observed on the
very first epoch. This ensures all observations are carried out as near differentially
as possible.

\begin{figure*}
   \centering\includegraphics[width=70mm]{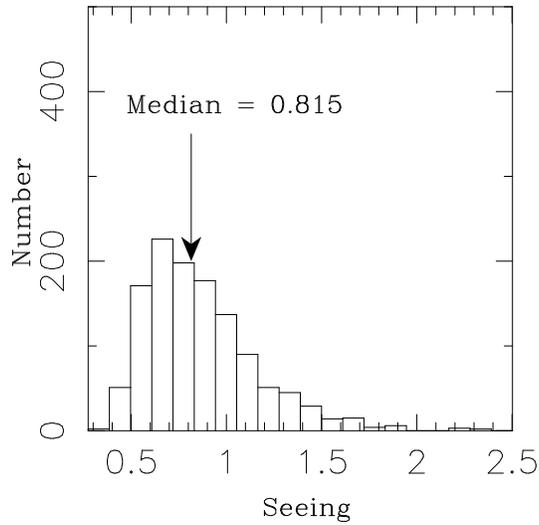}
   \caption{Seeing histogram for all observations in this program.}
   \label{seeing}
\end{figure*}

\begin{figure*}
   \centering\includegraphics[width=70mm]{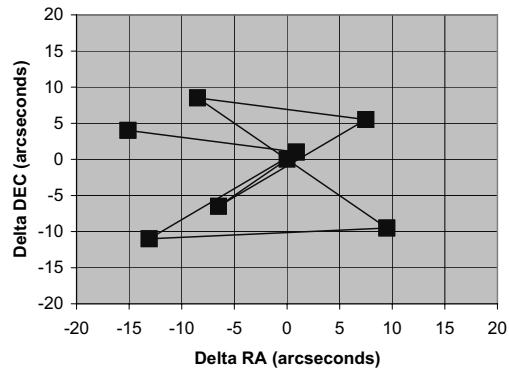}
   \caption{Fixed dither pattern used for all epoch observations. Offsets are
            in arcseconds from the initial base pointing position.}
   \label{jitter}
\end{figure*}

Seeing conditions over the course of this program varied. Figure \ref{seeing} shows
a histogram of the seeing full-width at half maxmimum for all our astrometric
observations. The median seeing was 0.82\arcsec, with 80\% of data being acquired
in seeing conditions between 0.55 and 1.25\arcsec.

In addition to these epoch observations, all
targets were also observed as they rose and set, so that DCR
calibrations for each target could be developed, following the technique
described in \citet{t95,t96}.  The J filter was chosen for these observations
as it offers the best contrast between sky- and T dwarf brightness. Typical 
near-infrared sky colours at La Silla are J--H=1.7, H--K=1.0-2.0 (2.0 for dark, 1.0 for bright)
(ESO SOFI on-line documentation). By contrast typical T dwarf colours are -0.5$<$J--H$<$0.5 and
H--K$<$0.5 \citep{bu2002a}, which means most T dwarfs are around a
magnitude brighter compared to the sky in J, than they are in H or K.

The sample of objects observed is listed in Table \ref{sample}, along with indications as to which
targets were observed at which epochs. SOFI has a nominal gain of 5.6e-/adu, and a nominal
read-noise of 14e- per exposure.

\begin{table*}
  \footnotesize
  \caption{NTT Infrared Parallax Target Sample.\label{sample}}
  \begin{tabular}{@{}lccccccccl@{}}
  \tableline\tableline
   Object&     Position                                                   & Apr & Jul & Mar & Apr & Jul & Mar & May & Ref.\\
         &    (J2000)                                                     & `00 & `00 & `01 & `01 & `01 & `02 & `02 &\\
  \tableline
\objectname[]{2MASS J05591914-1404488}   & 05\hh59\mm19\fs1$-$14\degr04\arcmin49\arcsec&  x  &     &     &  x  &     &  x  &&\tablenotemark{a}\\
\objectname[]{SDSS J10210969-0304201}    & 10\hh21\mm09\fs7$-$03\degr04\arcmin20\arcsec&     &  x  &  x  &  x  &  x  &  x  &&\tablenotemark{b}\\
\objectname[]{2MASS J10475385+2124234}   & 10\hh47\mm53\fs8$+$21\degr24\arcmin23\arcsec&  x  &  x  &  x  &  x  &     &  x  &&\tablenotemark{c}\\
\objectname[]{2MASS J12171110-0311131}   & 12\hh17\mm11\fs1$-$03\degr11\arcmin13\arcsec&  x  &  x  &  x  &  x  &  x  &  x  &&\tablenotemark{c}\\
\objectname[]{2MASS J12255432-2739466AB} & 12\hh25\mm54\fs3$-$27\degr39\arcmin47\arcsec&  x  &  x  &  x  &  x  &  x  &  x  &&\tablenotemark{c}\quad\tablenotemark{d}\\
\objectname[]{SDSS J12545390-0122474}    & 12\hh54\mm53\fs9$-$01\degr22\arcmin47\arcsec&     &  x  &  x  &     &  x  &  x  &&\tablenotemark{b}\\
\objectname[]{SDSS J13464645-0031504}    & 13\hh46\mm46\fs4$-$00\degr31\arcmin50\arcsec&  x  &  x  &  x  &  x  &  x  &  x  &&\tablenotemark{e}\\
\objectname[]{2MASS J15344984-2952274AB} & 15\hh34\mm49\fs8$-$29\degr52\arcmin27\arcsec&  x  &  x  &  x  &  x  &  x  &  x  &&\tablenotemark{f}\quad\tablenotemark{d}\\
\objectname[]{2MASS J15462718-3325111}   & 15\hh46\mm27\fs2$-$33\degr25\arcmin11\arcsec&  x  &  x  &  x  &  x  &  x  &  x  &&\tablenotemark{c}\\
\objectname[]{SDSS J16241437+0029156}    & 16\hh24\mm14\fs4$+$00\degr29\arcmin16\arcsec&  x  &  x  &  x  &  x  &  x  &  x  &&\tablenotemark{g}\\
  \tableline
\end{tabular}
\tablenotetext{a}{\citet{bu2000b}}
\tablenotetext{b}{\citet{le2000}}
\tablenotetext{c}{\citet{bu1999}}
\tablenotetext{d}{Shown to be a roughly equal binary by \citet{bu2003a}}
\tablenotetext{e}{\citet{ts2000}}
\tablenotetext{f}{\citet{bu2002a}}
\tablenotetext{g}{\citet{st1999}}
\end{table*}

%
%

\subsection{Object Names}

With the exception of $\epsilon$\,Ind\,B \citep{scholz2003}, all the objects
discussed in this paper have been discovered by either the 2MASS ({\tt www.ipac.caltech.edu/2mass}),
SDSS ({\tt www.sdss.org}) or DENIS ({\tt cdsweb.u-strasbg.fr/denis.html}) sky surveys, 
and have been given object names
by those surveys, based on their positions in J2000 coordinates. These names
have the advantage of being very specific and informative, and the disadvantage
of being lengthy and clumsy. Throughout this paper, therefore, we will generally
give an object's complete name when it is first used, and thereafter refer
to it (when not confusing to do so) by a shortened 2Mhhmm, SDhhmm or Dhhmm form
where hh and mm are the right ascension hour and minute components of its name.

\section{Analysis}

The analysis adopted for these data falls into two main areas: processing to produce linearised, flattened
and sky subtracted images, which was quite specific to the SOFI instrument; and astrometric processing of these
images, which identically follows that described in \citet{t93,t95,t96}.

\subsection{Processing SOFI data}

\begin{figure*}
   \centering\includegraphics[width=180mm]{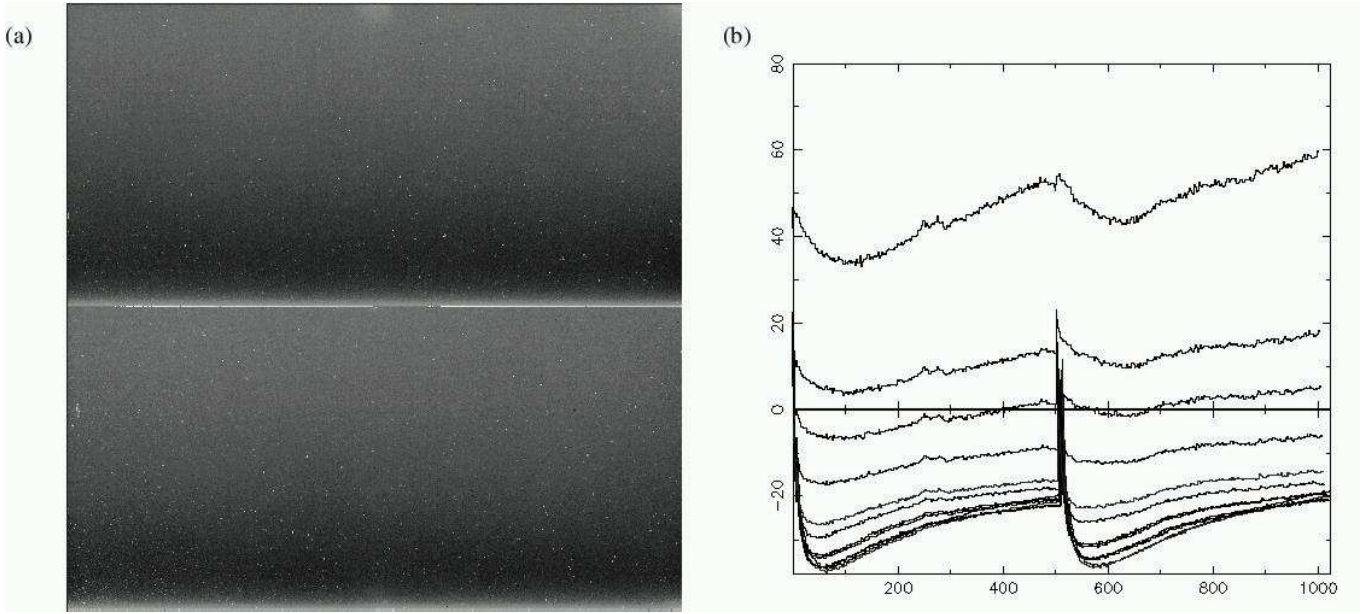}
   \caption{(a) Example of the SOFI shade pattern in a dark frame. (b) SOFI shade profiles obtained
   as vertical cuts through dark regions of the partially illuminated detector, for a range
   of count rates in the illuminated regions of the detector. The profiles span count rates
   in adu/pix from top to bottom of 3199, 2221, 1861, 1467, 1086, 958
   701, 685, 554, 483, 226 and 87.}
   \label{shade}
\end{figure*}

{\bf Dark frames and zero-points} : Dark frames obtained with SOFI reveal significant structure, 
which can be broken down into a few components.
\begin{enumerate}
  \item a significant ($\sim$50-100\,adu peak-to-peak) vertical structure, known as the ``shade'',
        which varies in intensity and shape with the
        overall level of illumination of the array;
  \item a small (1-20\,adu) dark current from the instrument and a small readout amplifier glow in each quadrant; and
  \item a tiny ($<$1\,adu) but fixed ``ray'' pattern left after the previous two are modeled and removed from
        dark current data.
\end{enumerate}
The shade pattern is of most concern as the remaining fixed patterns are small compared to the sky brightness.
Figure \ref{shade} shows a dark current image displaying the shade effect, together with a set of
vertical medianned profiles through the shade. Unfortunately, this shade profile is not constant - it 
varies with the intensity of
the overall level of illumination of the array during an exposure, meaning one has an unknown zero-point
for every pixel in every exposure.

Calibration of the shade was achieved as follows. A small aperture (used to mask the instrument entrance
when observing with one of the smaller fields of view) was inserted into the NTT focal plane, and
a series of flat fields obtained with varying exposure times. Because only the central quarter
of the array is illuminated by this procedure, it is possible to extract a shade profile from the edge
of each image. It is also possible to record the level of illumination of the array which produced
that shade profile. By performing a least-squares cubic polynomial fit through each pixel of these
shade profiles (which correspond to rows on the detector) as a function of array illumination, it is
possible to develop a parametrization for the shade profile. Using this parametrization it is
a simple matter to produce a shade profile estimate for each data image, and subtract it. The result
is an image with zero-point constant across the array.\footnote{Sample parameterizations
can be found at {\tt http://www.aao.gov.au/local/www/cgt/sofi}.}

\begin{figure*}
   \centering\includegraphics[width=180mm]{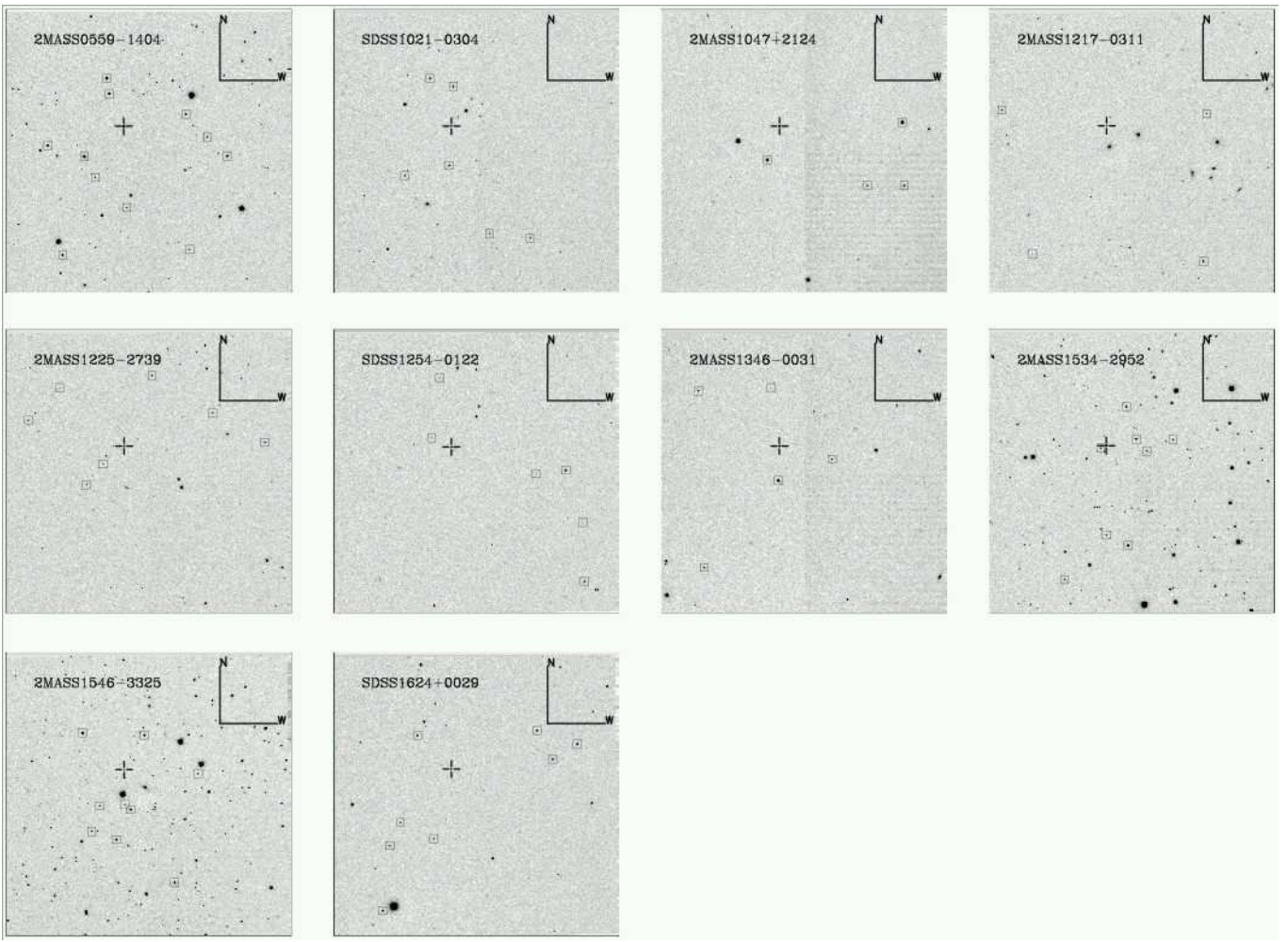}
   \caption{J-band finding charts with our target T dwarfs {\em (cross hairs)}, 
   and astrometric reference stars {\em (boxes)} highlighted. Each field
   is 295.2\arcsec\ on a side. The direction bars in each field are 1\arcmin\
   in size.}
   \label{charts}
\end{figure*}

%

\begin{figure*}
   \centering\includegraphics[width=80mm]{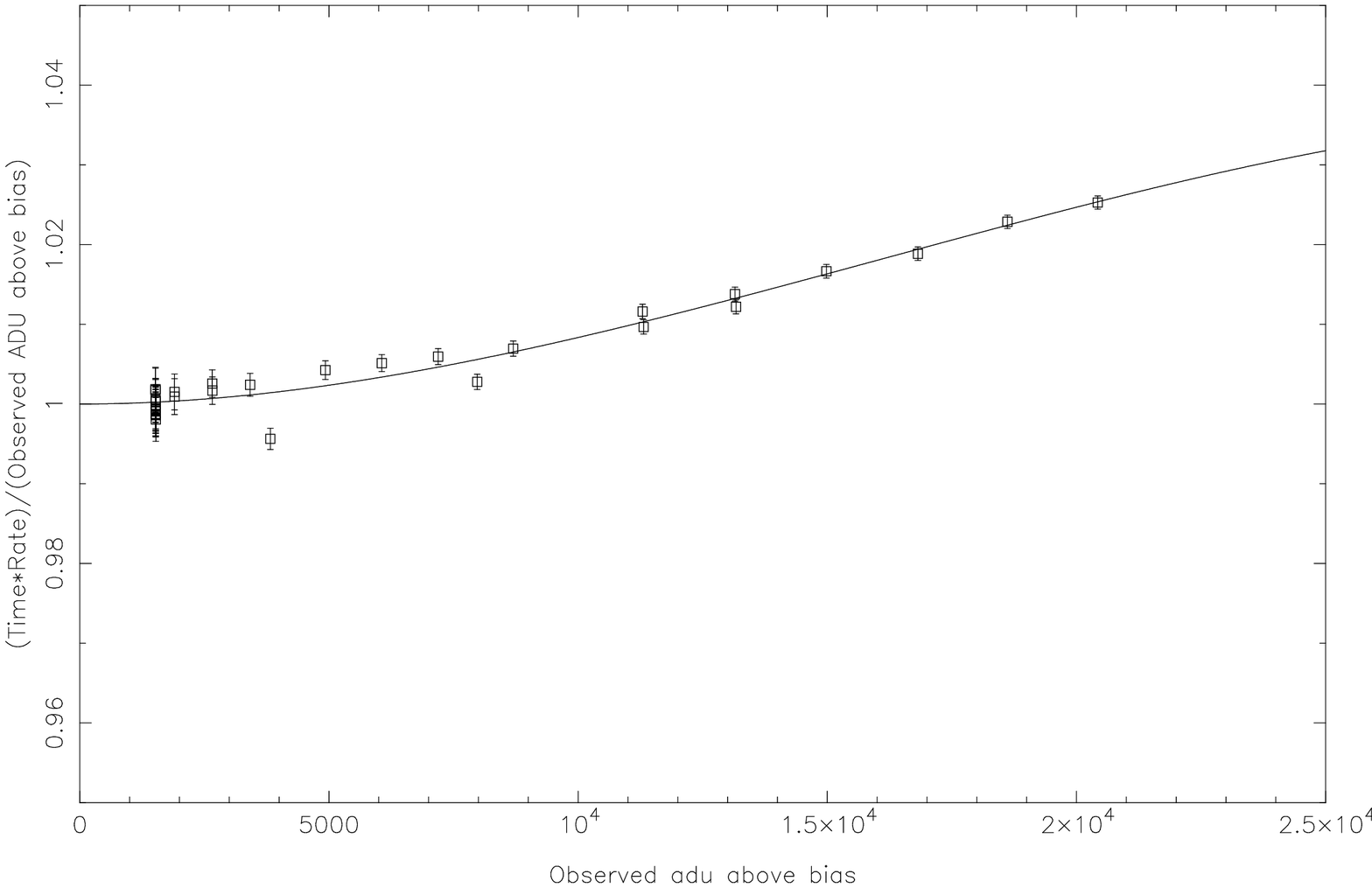}
   \caption{A linearity fit for a single SOFI quadrant. The data points plotted are
   the counts a perfectly linear detector would see, divided by the observed counts,
   as a function of observed counts. The fit is a fourth order polynomial, which 
   models the amount of non-linearity as a function of observed counts. 
   The actual linearity correction coefficients used are given in the text.}
   \label{samplelin}
\end{figure*}

{\bf Linearity} : All infrared detectors are non-linear to some extent. 
For SOFI, ESO usually recommends keeping
sky and target object intensities below 10,000\,adu in order to maintain linearity at better than 1\%.
Unfortunately, such an observing strategy is not useful for astrometry, which demands the largest
possible dynamic range to ensure targets (and reference stars) of widely differing magnitudes
are usable in widely varying seeing conditions. We therefore calibrated the linearity of SOFI
using the same shade profile data obtained above. Once the data have been shade corrected, they can
then be used to examine the response of each pixel to a constant light source over widely varying exposure
times. Repetition of a 'calibration' exposure time throughout the sequence allows the lamp's constancy
to be calibrated -- usually to within $\pm$ 0.5\%. A sample of the resulting 
linearity correction is shown in Fig. \ref{samplelin} for one of the SOFI quadrants. 
In all cases these tests were performed  independently for each quadrant, and 
the results were always consistent with the same linearity 
correction for all quadrants. A single correction was therefore derived as the mean of those in
each quadrant. Fig. \ref{samplelin}  shows that the detector is $\approx$2.5\% non-linear at 20,000adu above bias,
{\em but} that data can be obtained and linearity corrected even up to 25,000adu.

To linearise a pixel, then, with raw intensity $I_{ij}$, it is simply necessary to multiply it by the
polynomial P($I_{ij}$) = $a_0 + a_1\,I_{ij} + a_2\,I_{ij}^2 + a_3\,I_{ij}^3$. The coefficients adopted 
were $ a_0=1.0,a_1=0.0, a_2=1.11329\times10^{-10}, a_3=-2.46799\times10^{-15}$ 
for 2000 April - 2001 April, and 
$ a_0=1.0,a_1=0.0, a_2=8.6124661\times10^{-11}, a_3=-1.6986849\times10^{-15}$ for 2001 July - 2002 May.

{\bf Inter-quadrant Row Crosstalk} : HgCdTe detectors typically show an effect known as 
inter-quadrant row crosstalk \citep{xtalk}. This has the effect that a constant, small fraction 
of the total flux seen in each row is seen as crosstalk at the same row in all the other quadrants. 
Correction of this effect is straightforward. The detector is integrated up into a single
vertical column, then the two halves of this cut (Y=1-512 and Y=513-1024) are averaged,
multiplied by a single cross-talk constant, and subtracted from every column of the detector.
We found a crosstalk coefficient of 2.8$\times$10$^{-5}$ worked well.

{\bf Flat-fielding } : Flat-fielding was performed using dome flats. Because of the (variable) 
shade pattern present in every dome flat, NTT staff have developed an observing recipe to obtain
a ``special'' flat-field without a shade pattern present. Or one can used the shade calibration
procedure described above to correct standard dome-flat fields. 
Both were tried for this program and both provided
similar results. In the end every run's data was flattened with a ``special'' dome flat, 
as per usual for SOFI observing.

{\bf Sky subtraction } : Each group of eight 120s dithered exposures was then used to create
a normalised and medianned sky frame, which was re-normalised to each of the twelve observations
to perform sky subtraction. So that the data frames would maintain approximate photon-counting
errors, an appropriate constant sky level was then added back in to each frame

{\bf Astrometric processing} : Following this processing then, we have eight bias-subtracted, linearised, cross-talk corrected, flattened
and sky-subtracted data frames for each astrometric epoch. These were then subject to further
processing (ie. object finding and point-spread function fitting using DAOPHOT, DCR calibration
and proper-motion and parallax solution fitting) as eight individual
observations, in a manner  identical to that described in \citet{t93,t95,t96}.

\subsection{Astrometric Calibration of SOFI}
\label{scale}

Astrometric calibration observations were acquired in USNO Astrometric Calibration Region
M \citep{stone1999} on 2001 July 12 and 13. 
These consisted of sixteen 60s exposures (on each night) 
scattered throughout the 3.2\degr$\times$7.6\degr\
region which \citet{stone1999} have astrometrically calibrated.
These were processed identically to our main astrometric targets.
Reference catalogue positions were extracted from the USNO ACR catalogue\footnote{This data can be obtained from the Vizier service.}
 in SOFI-field-sized
regions around each nominal telescope pointing position. These positions were
then tangent projected (using the SLALIB library \citealt{sun67}) to provide reference data sets in
arcsecond offsets on the sky for each observation. These were matched
against the observed data to derive a set of linear (ie shift, scale and rotate)
transformations from the SOFI pixel positions to arcsecond offsets on the sky\footnote{This step
made extensive use of M.Richmond's excellent {\tt match} implementation of
the \citet{valdes95} object list matching algorithm, which is available at 
{\tt http://acd188a-005.rit.edu/match/}}.
These transformations determine the SOFI plate scale on this night to be
0.28826$\pm$0.00003\arcsec/pixel, and that the detector pixel's misalignment
with N-S (0.030$\pm$0.003\degr).
These data were also analysed to examine the amount of astrometric distortion (ie.
variability in the instrument plate scale with position in the field) present
in the SOFI optical train. The astrometric calibration data show there is no significant
astrometric distortion in the SOFI Large Field optics. The plate scale in the field
corners is the same as that in the field center to within 0.1\%. The SOFI field 
can be considered astrometrically flat to 0.1\%.

\begin{figure*} 
   \begin{center}
      \includegraphics[width=150mm]{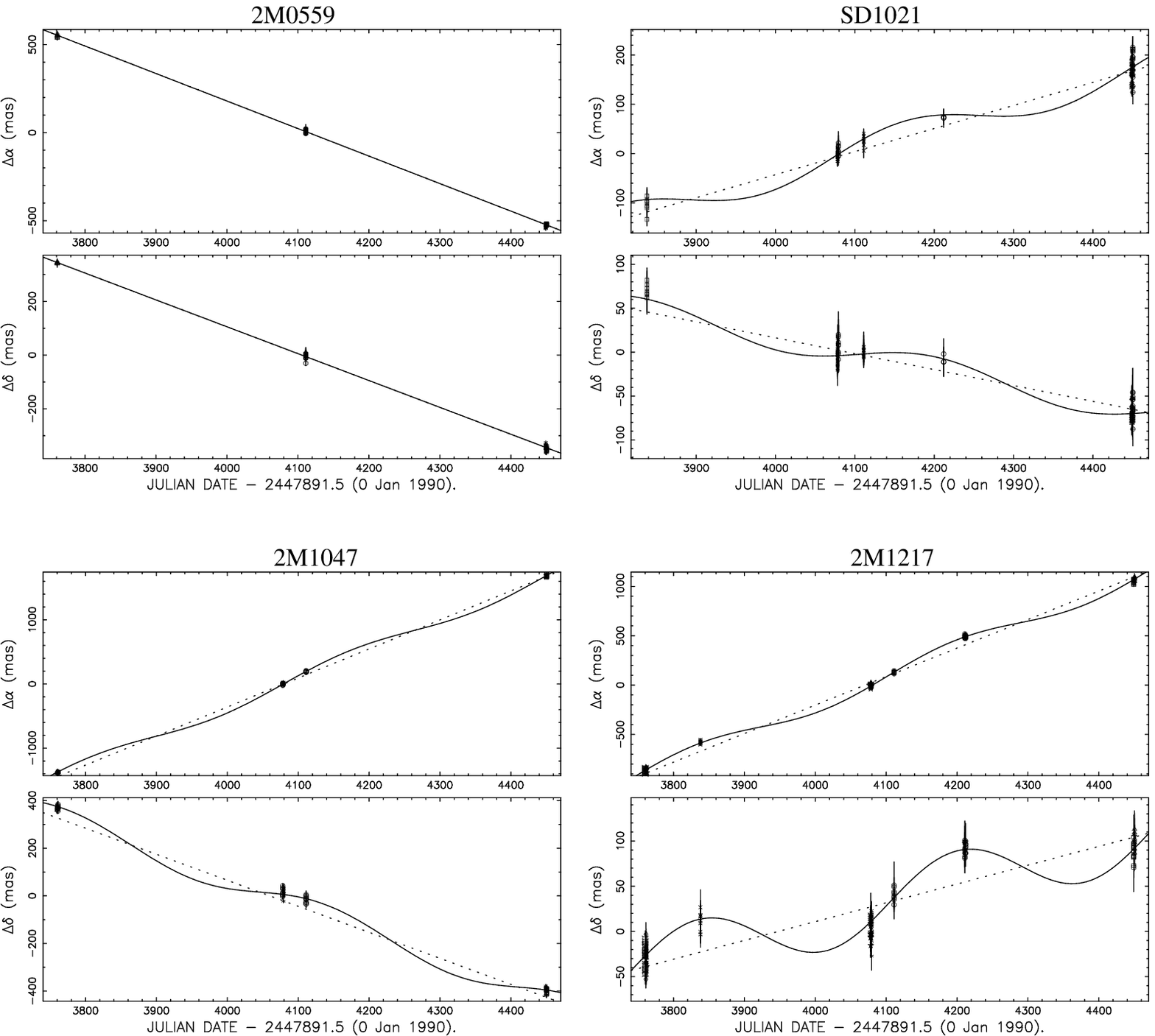}
   \end{center}
   \caption{Plots of the observed astrometric motions of our target T dwarfs, relative to the
   ensemble of selected background reference stars. For each object we show plots of their motion
   in right ascension ($\Delta\alpha$) and declination($\Delta\delta$) as a function of date.
   Also shown are the fitted proper motion + parallax (solid lines) and where a significant
   parallax has been measured, the fitted proper motion alone (dotted lines). The parameters
   fitted are given in Table \ref{results}.}
   \label{plots}
\end{figure*}

\begin{figure*}
   \figurenum{\ref{plots} {\em (cont.)}}
   \begin{center}
      \includegraphics[width=150mm]{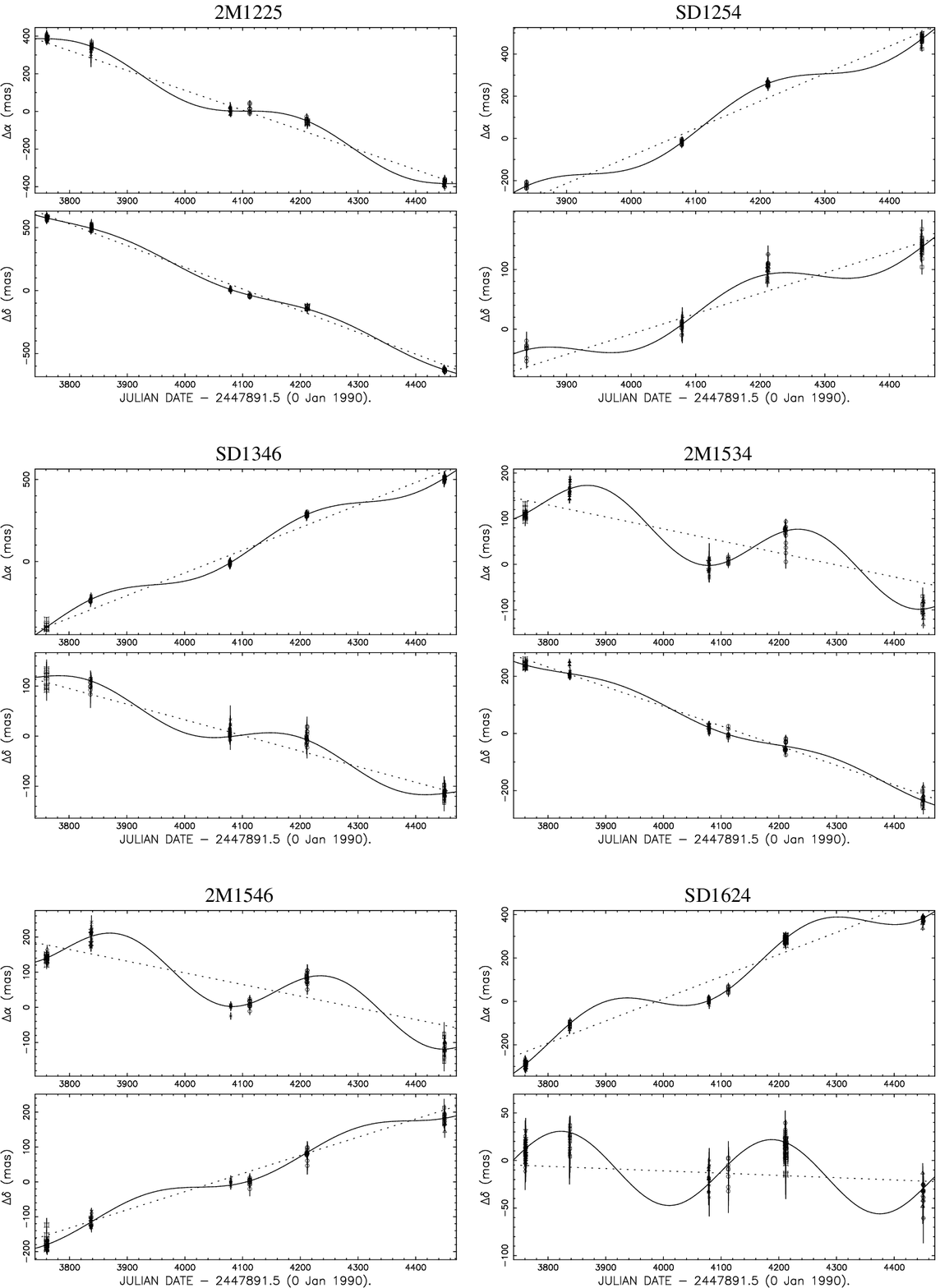}
   \end{center}
   \caption{}
   \label{plots2}
\end{figure*}

\begin{table*}
 \centering
 \begin{minipage}{100mm}
  \caption{NTT Infrared Parallax Sample - Relative Astrometry\label{results}}
  \begin{tabular}{@{}lrrrrr@{}}
  \tableline\tableline
   Object\footnote{See Table 1 for full object names}
         &N$_{f,n,s}$&$\pi_{r}$~~~ &$\mu_{r}$~~~&  $\theta_{r}$~~~~& V$_{tan}$ \\
         &         & (mas)       &  (mas)        & (\degr)~~~~   &(km/s) \\
  \tableline           
 2M0559  & 71,5,11 & (96.9)\footnote{\citet{da2002}}& 677.4$\pm$2.5 & 122.6$\pm$0.1 & 33.1$\pm$0.1\\
  S1021  &112,7,6  & 34.4$\pm$4.6& 183.2$\pm$3.4 & 248.8$\pm$1.0 & 25.2$\pm$2.4\\
 2M1047  & 70,7,4  &110.8$\pm$6.6&1698.9$\pm$2.5 & 256.4$\pm$0.1 & 72.7$\pm$4.4\\
 2M1217  &143,10,4 & 90.8$\pm$2.2&1057.1$\pm$1.7 & 274.1$\pm$0.1 & 55.2$\pm$1.4\\
 2M1225AB&128,12,7 & 75.1$\pm$2.5& 736.8$\pm$2.9 & 148.5$\pm$0.1 & 46.5$\pm$1.7\\
  S1254  &104,7,6  & 73.2$\pm$1.9& 491.0$\pm$2.5 & 284.7$\pm$0.1 & 31.8$\pm$1.0\\
 2M1346  &118,9,5  & 68.3$\pm$2.3& 516.0$\pm$3.3 & 257.2$\pm$0.2 & 35.8$\pm$1.4\\
 2M1534AB&140,11,8 & 73.6$\pm$1.2& 268.8$\pm$1.9 & 159.1$\pm$0.1 & 17.3$\pm$0.4\\
 2M1546  &150,10,9 & 88.0$\pm$1.9& 225.4$\pm$2.2 &  32.5$\pm$0.6 & 12.1$\pm$0.4\\
  S1624  &152,11,8 & 90.9$\pm$1.2& 373.0$\pm$1.6 & 268.6$\pm$0.3 & 19.5$\pm$0.3\\
\tableline                                    
\end{tabular}                             
\end{minipage}                            
\end{table*}                              

%
%

\section{Results for T dwarfs}

Astrometric solutions for our T dwarf targets were evaluated in a manner identical to that used
by \citet{t95}. Briefly the procedure is to transform (using a
linear transformation with rotation and a scale factor) all the
frames for a given object, onto a chosen master frame of good seeing (known
to have the detector rows and columns aligned with
the cardinal directions to within $\pm 0.1$\degr), using
a set of well exposed reference stars which were required to appear in every
frame; differential colour refraction (DCR -- see \citet{t95}) coefficients 
were then evaluated for each of these reference stars
(relative to the unknown `mean' DCR coefficient for the reference star set), and the
reference frame corrected for DCR; each frame was then re-transformed onto the
`master' frame; the DCR coefficient for the program object (relative to
the DCR-corrected reference frame) was then evaluated; the program object was
DCR corrected; and finally,  an astrometric solution in parallax and proper motion was
made independently for both the $\alpha$ and $\delta$ directions, using a
linear weighted-least-squares fit. Uncertainties arising from the DCR correction
and the residuals about the
reference frame transformation for each frame were carried through to this
solution fit, so observations taken in poor seeing or with poor
signal-to-noise due to cloud autmoatically receive low weight. The final
parallax was taken to be the weighted mean of the $\alpha$ and $\delta$
solutions. Finding charts for our target stars (taken from our SOFI data)
showing both the target and reference stars adopted can be seen in Fig. \ref{charts}.

The resulting relative astrometry is presented in Table \ref{results}, the
columns of which show; 
the number of frames (N$_f$), nights (N$_n$) and reference stars (N$_s$) used
in each solution; the parallax and proper motion solutions (relative to the
background reference stars chosen); and the derived tangential velocity for each
target )based on the measured parallax, except for 2MASS0559, for which we adopt
the distance of \citealt{da2002}). Plots of
these fits are shown in Fig. \ref{plots}. The reference stars used to obtain
this relative astrometry are typically within $\pm$1\,mag. of the apparent magnitude
of our target T dwarf. At these magnitudes (J=15-18) the reference stars will
most commonly be G- to early M type stars at distances of 500-2000pc. Thus although
we do not have the photometry available to estimate detailed corrections from
relative to absolute parallax, we can estimate with some confidence, that such
corrections will generally be less than 1\,mas in size, and so not significant 
in comparison
to our random astrometric uncertainties.

It is instructive to examine the root-mean-square residuals obtained for the
reference frame stars in our astrometry, since they tell us how
precise we can expect the astrometry of our target objects to be. Over the course of our program
we found that for a single 120s exposure, the median value of this
rms residual was 0.042\,pixel or 12.1\,mas, with 80\% of observations
having an rms residual between 6.9 and 20.2\,mas. Recall that at each epoch
we acquired 8 such observations in a total exposure time of 960s, which would
suggest the median precision from a single epoch is 12.1/sqrt(8) = 4.3\,mas.
Residuals within these groups of eight were somewhat correlated (presumably
because they are largely acquired in similar seeing conditions).

The USNO have published astrometry for three T dwarfs: 
2MASS0559, SDSS1254 \& SDSS1624 \citep{da2002}. While all three were
included in our program, insufficient epochs were obtained for 2MASS0559 
to measure a parallax. 
The equivalent relative parallax solution quantities for those we obtained (Table \ref{results}),
are given by Dahn et al. as;
for SDSS1254-01:       84.1$\pm$1.9\,mas,       496.1$\pm$1.8\,mas/y,                285.2$\pm$0.4\degr,
and for SDSS1624+00 :  90.7$\pm$2.3\,mas,       383.2$\pm$1.9\,mas/y,                269.6$\pm$0.5\degr.
These independent observations and solutions agree within uncertainties for almost all parameters --
the exception being the parallax for SDSS1254, for which the two
solutions are different by about 5-$\sigma$, though Dahn et al. do
comment that with only 1.2 years of data on this target their solution is
only considered to be preliminary.

Finally we note that with only 3 epochs of observation per year over two-plus years,
there is always the possibility that systematic errors on individual runs may
have impacted on our results. For example, a major change in SOFI's astrometric
distortion or a decollimation between the telescope and SOFI on a single run could
systematically effect our results. The only way to detect such problems is by detecting
a poor match between our astrometric model and the data we obtain, which is
difficult with less than 6 epochs. We believe the likelihood of this is small
because; (1) exactly the same automated telescope image analysis procedures were used
to control the NTT's primary figure throughout every night of every run, 
making the chance of an unusual NTT collimation with SOFI unlikely; (2) SOFI's astrometric distortion (as we have shown
above) is tiny, so changes in it can have only negligible effect; (3) SOFI is a Nasmyth
mounted instrument, and so is always mounted horizontally and subject only to rotation about
its optical axis, greatly reducing the likelihood of flexure within the instrument; and finally
(4) because infrared instruments sit in temperature-controlled dewars they suffer almost
none of the temperature-dependent flexure and defocus effects present in optical reimaging
systems, and they are also much less prone to being opened and modified over the course
of an astrometric program. On-going monitoring and independent observations
by independent programs are the best way to test for unforeseen systematic errors,
and we look forward to checking our results against programs being carried out elsewhere.

\section{Discussion}

\subsection{Differential Colour Refraction for T dwarfs}
\label{sec_dcr}

\begin{figure*}
   \centering\includegraphics[width=80mm]{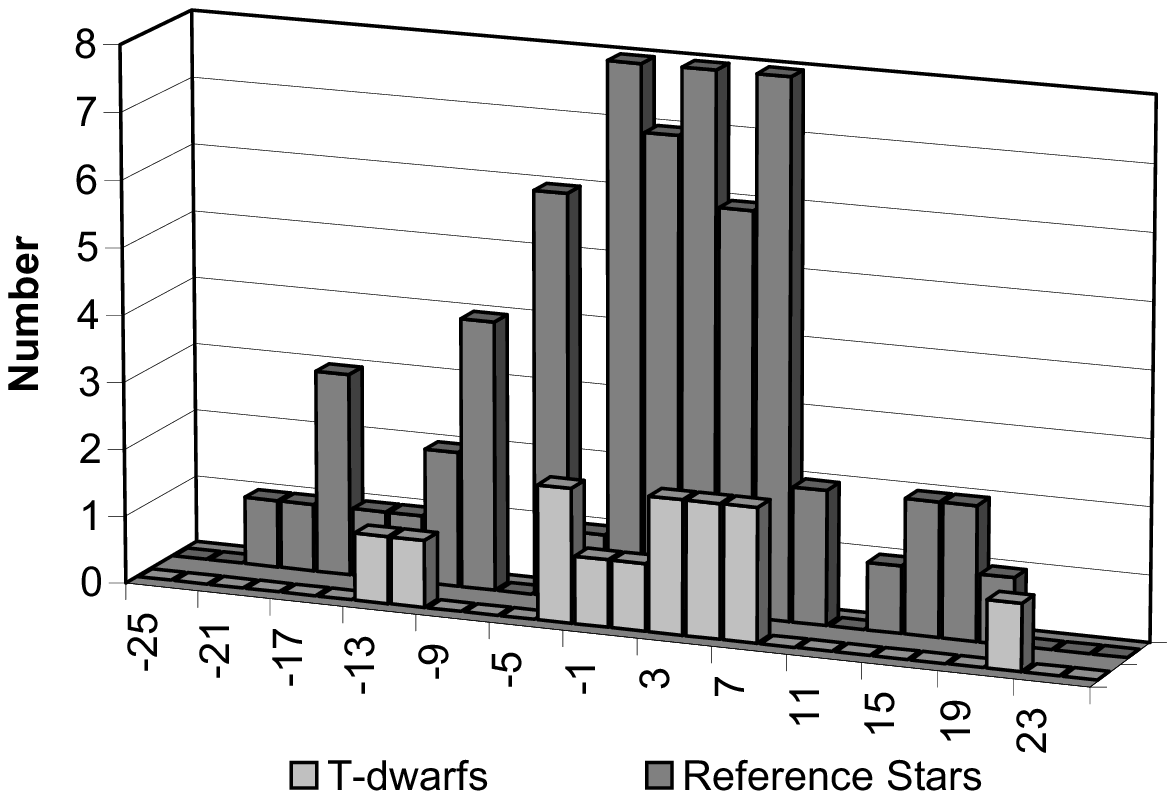}
   \caption{DCR coefficient histograms for reference stars and 
   programme T dwarfs, in units of mas/Tan(Zenith Angle). See Section \ref{sec_dcr} for more details.}
   \label{dcrhist}
\end{figure*}

An interesting result of the DCR calibrations we performed for our T dwarf
targets, was to find that T dwarfs have
effective wavelengths in the J-band which are essentially indistinguishable
from the ensemble of background reference stars against which their
positions are measured. This is shown in Fig. \ref{dcrhist} which plots histograms
of the DCR coefficients determined for reference stars and programme
T dwarfs - the similarities in the ensemble values are clear. 
(These coefficients were derived using the method described in 
\citealt{t93} and \citealt{t96}. Typical uncertainties in the individual determinations 
are $\approx \pm 2-6$\,mas$/$tan(ZA).) As a result, though we have calibrated and
applied DCR corrections to our data, such a procedure is
not strictly necessary for near-infrared observations of
T dwarfs. These
observations, therefore,  are {\em not} rigidly tied to being carried out near the
meridian, which adds enormously to the flexibility and efficiency of
infrared parallax programmes.

\subsection{Photometry for L  and T dwarfs}

There are currently only a few large and systematic photometric databases for late M, L 
and T dwarfs extant. The first is the photometry from the 2MASS database \citealt{2mass}, which
has the advantage of being a well established photometric system which covers the whole
sky, and includes all of our T dwarf targets, and almost of all of the other
known L  and T dwarfs \citep{burg_thesis}. Unfortunately, the photometry for these objects in the
J and K$_s$ 2MASS bands is often near the 2MASS photometric limits, so typical uncertainties
of $\pm$0.1\,mag. or greater are not uncommon. Moreover, because 2MASS does not include
an optical passband, colour information has to rely on the J--K colour,
which is typically small compared to the photometric precision, as well as giving only
a small wavelength ``lever arm'' on the spectral evolution of L  and T dwarfs. 

We make use of absolute M$_J$ and M$_{Ks}$
values on the 2MASS system compiled by \citet{burg_thesis} which is based
on the parallaxes presented in \citet{da2002}. \citet{da2002} also present
optical photometry in the $I_C$ passband, and J,H,K photometry in
a photometric system approximating that of the CIT system of \citet{elias1982},
as well as data form other work transformed onto this system.

A second extensive database is that compiled by \citet{le2002}. This includes Z band photometry
(on a UKIRT defined photometric system) as well as J,H,K photometry transformed by the
authors onto the MKO photometric system (see \citealt{le2002}, Section 3 for details). Because
these data were acquired with a 4m telescope, their photometric precision is much
higher than that for 2MASS. Great care should be taken in inter-comparing these two
sets of photometry -- the systematic differences between the two photometric systems
are {\em very} significant. This is {\em particularly} true of the Z photometric
system of \citet{le2002}, which is based on a relatively narrow interference
filter (0.851-1.056\,$\mu$m) used with a HgCdTe infrared array, leading to effective wavelengths
for L  and T dwarfs of $\approx$1.0\,$\mu$m, unlike the more common optical Z-type
observations which are based on long-pass filters ($\ga$0.85\,$\mu$m) and the declining sensitivity
of CCDs at $>$1\,$\mu$m, leading to effective wavelengths $\approx$0.9\,$\mu$m. 
\footnote{Indeed the two are so different that a distinctive
distinctive name -- Y -- for these HgCdTe-based Z magnitudes is being widely adopted}. UKIRT
Z photometry should not be assumed to be directly comparable with optically based Z
photometry. In the discussion that follows, therefore, we
will discuss only features in the absolute magnitudes {\em within an individual
photometric system}. For this reason we do not make use of the more heterogeneous J,H,K
compilation of \citet{da2002}.

To these data sets we add observations of the recently announced
T dwarf $\epsilon$\,Ind\,B \cite{scholz2003}, which has a mean I=16.7$\pm$0.1 and
2MASS photometry of J=11.91$\pm$0.04 and Ks=11.21$\pm$0.04 (Burgasser, priv.comm.).

\subsection{Photometric corrections for known binaries}
\label{bincor}

\begin{figure*}
   \centering\includegraphics[width=100mm]{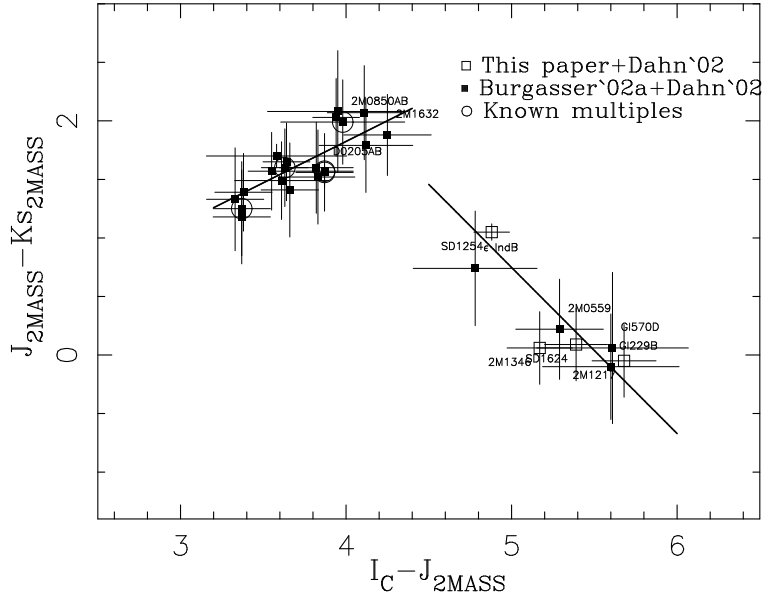}
   \caption{Plot of I$_C$--J against J--Ks for L  and T dwarfs from \citet{burg_thesis} 
   and \citet{da2002}. The data breaks into two clear sequences for the L  and T dwarfs.
   The solid lines are least-squares linear fits to the data as follows: 
   for L dwarfs J--Ks = 0.707(I$_C$--J) - 1.003 for 3.2$<$I$_C$--J$<$4.4 ($\sigma=0.14$);
   for T dwarfs J--Ks = 7.598 - 1.371(I$_C$--J) for 4.4$<$I$_C$--J$<$6.0 ($\sigma=0.25$).}
   \label{ij_jk}
\end{figure*}

\begin{table*}
 \centering
 \begin{minipage}{180mm}
  \footnotesize
  \caption{L  and T dwarf binary magnitudes\label{binaries}}
  \begin{tabular}{@{}lcccccl@{}}
  \tableline\tableline
System     &  Comp. & $\Delta$I$_C$\footnote{Difference in magnitude between the component
and the total magnitude of the system in this passband.}
 & $\Delta$J$^a$ & $\Delta$K$^a$ & Sp.T.\footnote{\citealt{bu2003a}} & 
 Notes\footnote{R01 - \citealt{reid2001}, K99 - \citealt{ko1999}, L01 - \citealt{le2001}, M99 - \citealt{mbb1999}, B03 - \citealt{bu2003a}} \\
  \tableline
\objectname[]{2MASS J0746425+200032AB}   & A & 0.50 & 0.54 & 0.59 & L0.5 & $\Delta$I$_C$,$\Delta$J R01. $\Delta$K Fig.\ref{ij_jk} \\
                                         & B & 1.12 & 1.01 & 0.95 & L0.5 & $\Delta$I$_C$,$\Delta$J R01. $\Delta$K Fig.\ref{ij_jk} \\
\objectname[]{2MASS J1146345+223053AB}   & A & 0.61 & 0.64 & 0.67 & L3 & $\Delta$I$_C$,$\Delta$J R01. $\Delta$K Fig.\ref{ij_jk} \\
                                         & B & 0.92 & 0.87 & 0.84 & L3 & $\Delta$I$_C$,$\Delta$J R01. $\Delta$K Fig.\ref{ij_jk} \\
\objectname[]{2MASSs J0850359+105716AB}  & A & 0.29 & 0.55 & 0.26 & L6 & $\Delta$I$_C$,$\Delta$J R01. $\Delta$K Fig.\ref{ij_jk} \\
                                         & B & 1.63 & 0.99 & 1.67 & T2:& $\Delta$I$_C$ R01. $\Delta$J Fig.\ref{cmd_iz}. $\Delta$K Fig.\ref{ij_jk} \\
\objectname[]{DENIS-P J0205290-115925AB}   & A & 0.75 & 0.75 & 0.75 & L7 & Assumed equal mass binary K99,L01 \\
                                         & B & 0.75 & 0.75 & 0.75 & L7 & Assumed equal mass binary K99,L01 \\
\objectname[]{DENIS-P J1228138-154711AB}   & A & 0.54 & 0.66 & 0.75 & L5 & $\Delta$J M99, $\Delta$K K99, $\Delta$I$_C$ Fig.\ref{ij_jk} \\
                                         & B & 1.02 & 0.86 & 0.75 & L5 & $\Delta$J M99, $\Delta$K K99, $\Delta$I$_C$ Fig.\ref{ij_jk} \\
\objectname[]{2MASS J12255432-2739466AB}  & A & 0.24 & 0.28 & 0.16 & T6 & $\Delta$I$_C$,$\Delta$J B03, $\Delta$K Fig.\ref{ij_jk} \\
                                         & B & 1.76 & 1.63 & 2.15 & T8 & $\Delta$I$_C$,$\Delta$J B03, $\Delta$K Fig.\ref{ij_jk} \\
\objectname[]{2MASS J15344984-2952274AB}  & A & 0.53 & 0.75 & 0.75 & T5.5 & $\Delta$I$_C$,$\Delta$J B03, $\Delta$K assumed 0.75 \\
                                         & B & 1.03 & 0.75 & 0.75 & T5.5 & $\Delta$I$_C$ B03, $\Delta$J,$\Delta$K assumed 0.75 \\
\tableline                                    
\end{tabular}                             
\end{minipage}                            
\end{table*}                              

%

Several of the systems published in the photometric compilations listed above
are known to be binaries, having been resolved either from the ground \citep{ko1999,le2001},
or using HST \citep{mbb1999,reid2001,bu2003a}. Unfortunately, not all these systems have 
measured magnitude differences in
all the passbands of interest, so we are forced to estimate magnitudes 
for the A and B components of these systems based on available
colour-colour relationship data. {\em In some cases (especially 2M0850B) these
extrapolations are large, and the decomposed magnitudes should be treated
as indicative only.} Table \ref{binaries}
shows the magnitude differences between each component and the {\em total} magnitude
of each system, along with estimated spectral types from \citet{bu2003a}. 
Because of the similarity between the effective wavelengths of
the HgCdTe-based Z and J bands,
we assume that the magnitude differences in Z are the same as those derived at J.
We do not differentiate here between UKIRT and 2MASS J and K bands.

{\em 2M0746AB, 2M1146AB \& 2M0850AB} were observed by \citet{reid2001} in
the HST F814W filter from which magnitude differences for the components were
derived in the I$_C$ passbands. Infrared J-band magnitude differences were
then estimated using the L dwarf sequence of the M$_{I}$ versus I$_C$--J 
colour-magnitude diagram (which has a roughly constant slope). 
With the exception of 2M0850B,
this procedure will be adequate for all the L dwarfs, and
those values are shown in Table \ref{binaries}. 2M0850B is an exception
because its absolute magnitude at
I$_C$ is so faint that it must be an early- to mid-T dwarf, rather than an L dwarf.
And, as we show in Section \ref{cmds}, the colour-magnitude diagram is not even
remotely linear across the L-T dwarf transition. For 2M0850B, therefore,
we have used the M$_I$ for the AB system of \cite{da2002},
and the magnitude differences of \citet{reid2001} to derive for the B component 
M$_I$ = 20.02$\pm$0.23. The colour-magnitude diagrams in Section \ref{cmds} then imply
I$_C$--J $\approx$ 5.1$\pm$0.2, from which we derive the B component J magnitude
difference shown in the Table.

To derive K-band magnitudes for the components of these systems, we have plotted 
J--Ks versus I$_C$--J for all the L  and T dwarfs in \citet{burg_thesis} and \citet{da2002} in
Figure \ref{ij_jk}. The data reveal two separate sequences -- the
L dwarfs in which J--Ks becomes redder along with I$_C$--J, and the
T dwarfs in which the reverse holds. The two lines on the plot are linear fits
to these two regimes (arbitrarily divided at I$_C$--J=4.4). From these relations
we predict J--Ks colours for L  and T dwarfs from their I--J colours, and so derive the
magnitude differences for each component in Table \ref{binaries}.

{\em D0205AB \& D1228AB} were both observed by \citet{ko1999} in the K band at
Keck, and D0205AB was independently observed at UKIRT by \citet{le2001}. 
D1228 was also observed in the J band with HST by \citet{mbb1999}. D0205 was
found to be a pair of objects with equal brightness at K, and in the absence of
any other information we assume it to be an equal mass binary. D1228 is a nearly
equal mass binary -- from the marginal J--K colour difference between the two 
components we can extrapolate to a magnitude difference between the components
at I of 0.48.

{\em 2M1225AB, 2M1534AB} have been observed by \citet{bu2003a} with HST in the F814W 
and F1042M filters (the latter enabling the derivation of approximate J magnitude
differences for the systems). Once again we use the I--J colours of these objects
to extrapolate to K magnitude differences for 2M1225AB's components. For 2M1534AB the magnitude differences
estimated at F1042 are only marginally different from zero, so we assume equal
brightnesses in this system at J and K.

\subsection{Spectral-Type-Magnitude relations for L  and T dwarfs}

\begin{table*}
 \centering
 \begin{minipage}{170mm}
 \scriptsize
  \caption{T dwarf Photometry \& Absolute Magnitudes for NTT Parallax Sample}
  \label{absolute}
  \begin{tabular}{@{}llccccccccccl@{}}
  \tableline\tableline
Object &  T\footnote{Spectral types on the \citet{bu2000a} system} 
                        &\multicolumn{2}{c}{2MASS\footnote{\citet{burg_thesis}. K$_s$ entries marked ``:'' are upper limits.}}
                                                                                                                &\multicolumn{3}{c}{MKO\footnote{\citet{le2002}, 
                                                                                                                                    except for 2M1534 which is Leggett, priv.comm. J,K uncertainties typically $\pm$0.03, Z--J typically $\pm$0.05 }}
                                                                              &\multicolumn{2}{c}{2MASS}        &\multicolumn{3}{c}{MKO}                           & \\
           &       &       J        &       K$_s$    &   J   &    K    &  Z-J &    M$_J$       &     M$_{Ks}$   & M$_{Z}$&     M$_J$      &    M$_K$       & \\
  \tableline
SD1021  & T3    & 16.26$\pm$0.10 & 15.10$\pm$0.18  & 15.88 &  15.26  & 1.78 & 13.94$\pm$0.29 & 12.78$\pm$0.33 & 15.34$\pm$0.27 & 13.56$\pm$0.27 & 12.94$\pm$0.27 & \\
2M1047  & T6.5  & 15.82$\pm$0.06 & 16.30$\pm$0.30: & 15.46 &  16.20  & 1.93 & 16.05$\pm$0.14 & 16.52$\pm$0.33 & 17.61$\pm$0.13 & 15.68$\pm$0.13 & 16.42$\pm$0.13 & \\
2M1217  & T7.5  & 15.85$\pm$0.07 & 15.90$\pm$0.30: & 15.56 &  15.92  & 2.00 & 15.64$\pm$0.09 & 15.69$\pm$0.30 & 17.35$\pm$0.06 & 15.35$\pm$0.06 & 15.71$\pm$0.06 & \\
2M1225  & T6    & 15.22$\pm$0.05 & 15.06$\pm$0.15  & 14.88 &  15.28  & 1.89 & 14.60$\pm$0.09 & 14.44$\pm$0.17 & 16.15$\pm$0.08 & 14.26$\pm$0.08 & 14.66$\pm$0.08 & \\
SD1254  & T2    & 14.88$\pm$0.04 & 13.83$\pm$0.06  & 14.66 &  13.84  & 1.74 & 14.20$\pm$0.07 & 13.15$\pm$0.08 & 15.72$\pm$0.06 & 13.98$\pm$0.06 & 13.16$\pm$0.06 & \\
2M1346  & T6    & 15.86$\pm$0.08 & 15.80$\pm$0.30: & 15.49 &  15.73  & 2.24 & 15.03$\pm$0.11 & 14.97$\pm$0.31 & 16.90$\pm$0.08 & 14.66$\pm$0.09 & 14.90$\pm$0.08 & \\
2M1534  & T5.5  & 14.90$\pm$0.04 & 14.86$\pm$0.11  & 14.60 &  14.91  &      & 14.24$\pm$0.05 & 14.19$\pm$0.12 &                & 13.94$\pm$0.05 & 14.25$\pm$0.05 & \\
2M1546  & T5.5  & 15.60$\pm$0.05 & 15.42$\pm$0.17  &       &         &      & 15.32$\pm$0.07 & 15.14$\pm$0.18 &                &                &                & \\
SD1624  & T6    & 15.49$\pm$0.06 & 15.40$\pm$0.30: & 15.20 &  15.61  & 2.12 & 15.28$\pm$0.07 & 15.19$\pm$0.30 & 17.11$\pm$0.04 & 14.99$\pm$0.04 & 15.40$\pm$0.06 & \\
\tableline                                    
\end{tabular}                             
\end{minipage}                            
\end{table*}                              

\begin{figure*}
   \centering\includegraphics[width=140mm]{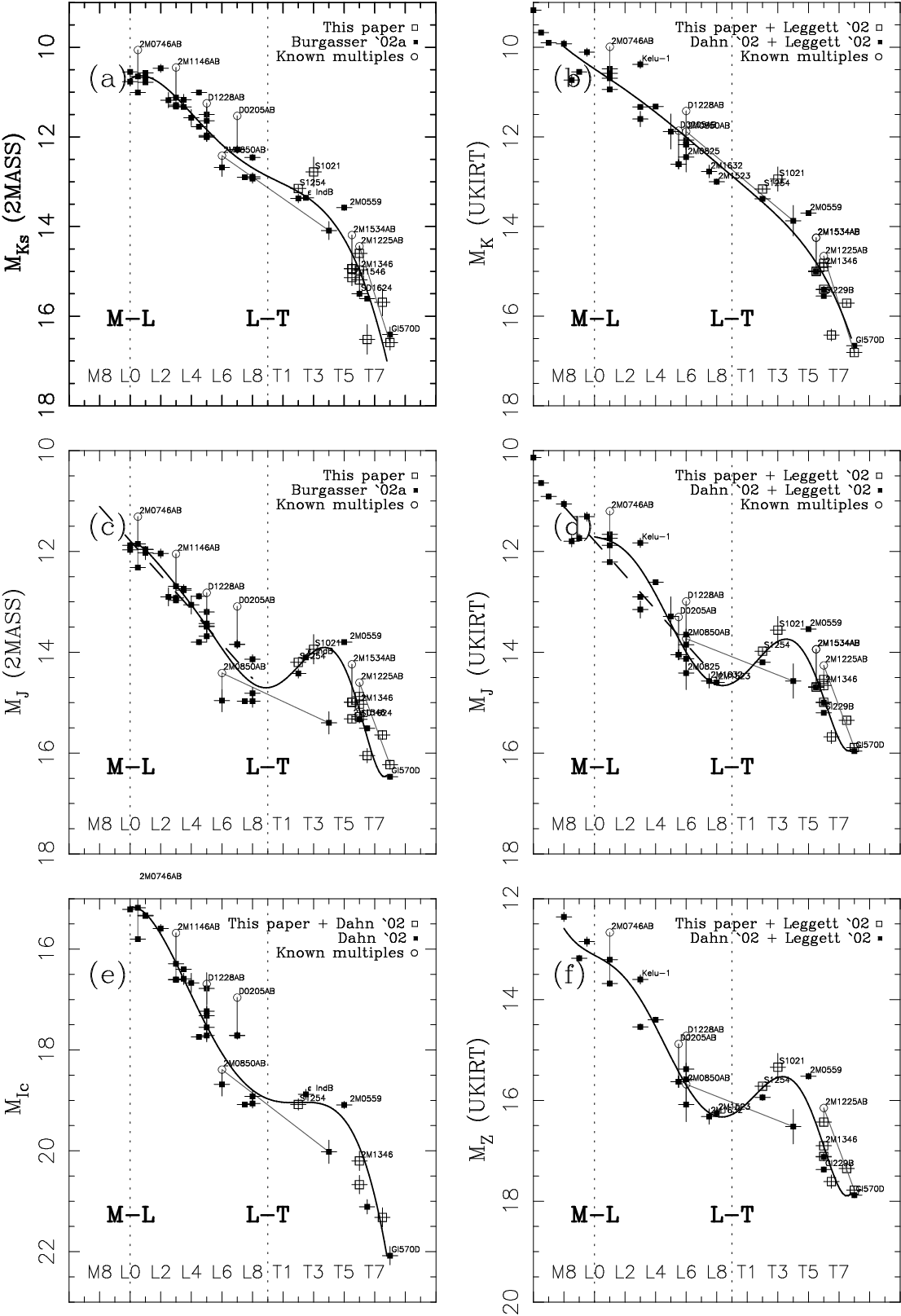}
   \caption{Absolute magnitudes in a variety of bandpasses, as a function of L and T spectral types on the
   Kirkpatrick et al. and Burgasser et al. systems. Panels (a) and (c) show absolute M$_J$ and M$_{Ks}$
   from 2MASS photometry in \citet{burg_thesis,scholz2003} and this paper. Panels (b),(d) and (f) show 
   absolute M$_J$, M$_K$ and M$_Z$ on the \citet{le2002} UKIRT system. Panel (e) shows M$_{Ic}$ from
   photometry in \citet{da2002}. All parallaxes are from either this work {\em (open squares)}
   or \citet{da2002} \& \citet{scholz2003} {\em (filled squares)}.
   The dotted line in panels (c) and (d) is the spectral-type-magnitude relation of \citet{da2002}.
   Known unresolved binaries are encircled, and decomposed into their resolved components as described in 
   Section \ref{bincor}.
   The solid heavy lines are seventh-order polynomial weighted fits to the data (see Section \ref{relations}
   for coefficients). The decomposed JK magnitudes and spectral type of 2M0850B are much more uncertain than the formal
   uncertainties plotted would indicate, and should be treated as indicative only.}
   \label{abs_spt}
\end{figure*}

%

Table \ref{absolute} lists \citet{burg_thesis}, \citet{da2002} and \citet{le2002} photometry for our NTT parallax
sample, along with the resulting absolute magnitudes in these systems. Also listed
are spectral types on the scheme of \citet{bu2002a}.

Figure \ref{abs_spt} shows plots of spectral type against M$_Z$, M$_I$, M$_J$ and M$_{Ks}$/M$_K$. 
Also shown are absolute magnitudes for late M and L dwarfs
using parallaxes and 2MASS photometry from \citet{da2002} for the 2MASS panels,
and parallaxes from \citet{da2002} and UKIRT photometry from \citet{le2002} for
the UKIRT panels. The spectral types are on the system of \citet{ki1999} for the
M and L dwarfs, and \citet{bu2002a} for the T dwarfs. Known multiple systems are noted with circles, 
and decomposed into their component magnitudes as discussed above.

The two K-band plots (Fig. \ref{abs_spt}a and \ref{abs_spt}b) indicate that in both
systems, the L-T transition is marked by a steepening of the spectral-type-magnitude
 relation. In general, however, the relationship between absolute magnitude at K
and spectral type is well behaved for the purpose of estimating absolute magnitudes from
spectral types.
This is certainly {\em not} true in the 2MASS and UKIRT J passbands 
(Fig. \ref{abs_spt}c and \ref{abs_spt}d). Indeed, both sets of data indicate
a strong inflection (a ``hump'') in the relationship between absolute magnitude and spectral
type for early T dwarfs -- as a class, the T0-T4 brown dwarfs
have absolute magnitudes {\em brighter} than the latest L dwarfs by a magnitude or more.
Put another way, a simple extrapolation of the spectral-type-magnitude relationship 
for L dwarfs (eg. that from \citet{da2002} shown in the figure) underestimates
the absolute magnitude of the early- to mid-T dwarfs by up to two magnitudes. 
This ``early T hump'' has been noted previously \citep{da2002},
though on the basis on fewer T dwarf parallaxes. It has been
suggested \citep{burg_thesis} that binarity could be the cause for  
 early T dwarfs being more luminous than
the late L dwarfs. While it is certainly true that the L  and T dwarfs which have
been resolved as binaries are displaced to apparently high
absolute magnitudes when plotted as unresolved objects, the addition of new parallaxes 
would seem to indicate the over-luminosity of early T dwarfs
is a general property, rather than being due to the selection of objects 
which happen to be binaries.
Moreover, the magnitude or more of over-luminosity is too large an effect 
to be due equal-mass binarity which can produce a brightening of only 0.75\,mag.
A similar (though possibly less pronounced) inflection is seen in the
M$_Z$ relation (Fig. \ref{abs_spt}f), while the  M$_I$ relation (Fig. \ref{abs_spt}e)
would appear to be almost as monotonic as that at K, though with a more pronounced
inflection at the L-T boundary.
Having said this, however, 2M0559 continues to appear to be over-luminous compared
to the other early- to mid-T dwarfs in the figure. \citet{bu2003a} failed to resolve
a binary companion in this system with HST, implying that if it is a binary it must have
a separation of less than 0.5\,a.u.
We also note that it has been suggested \citep{tsuji2003} that the selection of preferentially
young objects could produce the ``early T hump'' -- we discuss this further in Section \ref{sec_hump}.

There are good physical reasons for expecting a monotonic relationship between
effective temperature (T$_{\mathrm eff}$) and luminosity (L) in these objects,
since these quantities are directly determined by interior (rather than photospheric)
properties. However, it must be remembered that as proxies for T$_{\mathrm eff}$
and luminosity, absolute magnitude in a given passband and spectral type are far from perfect.
Spectral typing is in essence an arbitrary allocation of a quantity to an object based on what
its spectra look like -- there is no guarantee that the relationship between
spectral type and T$_{\mathrm eff}$ (even if monotonic) should not have significant
changes in slope. Similarly, the relationship between absolute magnitude in a given passband
and luminosity is even more problematic. From the spectra of objects ranging
from L to T spectral types, and indeed from their J--K colours \citep{da2002}, 
we know that {\em significant} changes take place in their photospheres. There is
significant redistribution of flux in the spectra of brown dwarfs
across the L-T transition. We should not be surprised if this results in the relationship between luminosity
absolute magnitude in a given passband not only containing changes in slope, but not even being monotonic.

Given our current parallax database, spectral type is a 
very poor proxy for absolute magnitude in the Z and J bands from mid-L to mid-T spectral types.
The sequences in Fig.\ref{abs_spt} will need to be filled in by many 
more L  and T dwarfs before precise absolute magnitudes can be estimated from 
spectral types with confidence.

\subsection{Colour Magnitude Diagrams for L  and T  dwarfs}
\label{cmds}

\begin{figure*}
   \centering\includegraphics[width=170mm]{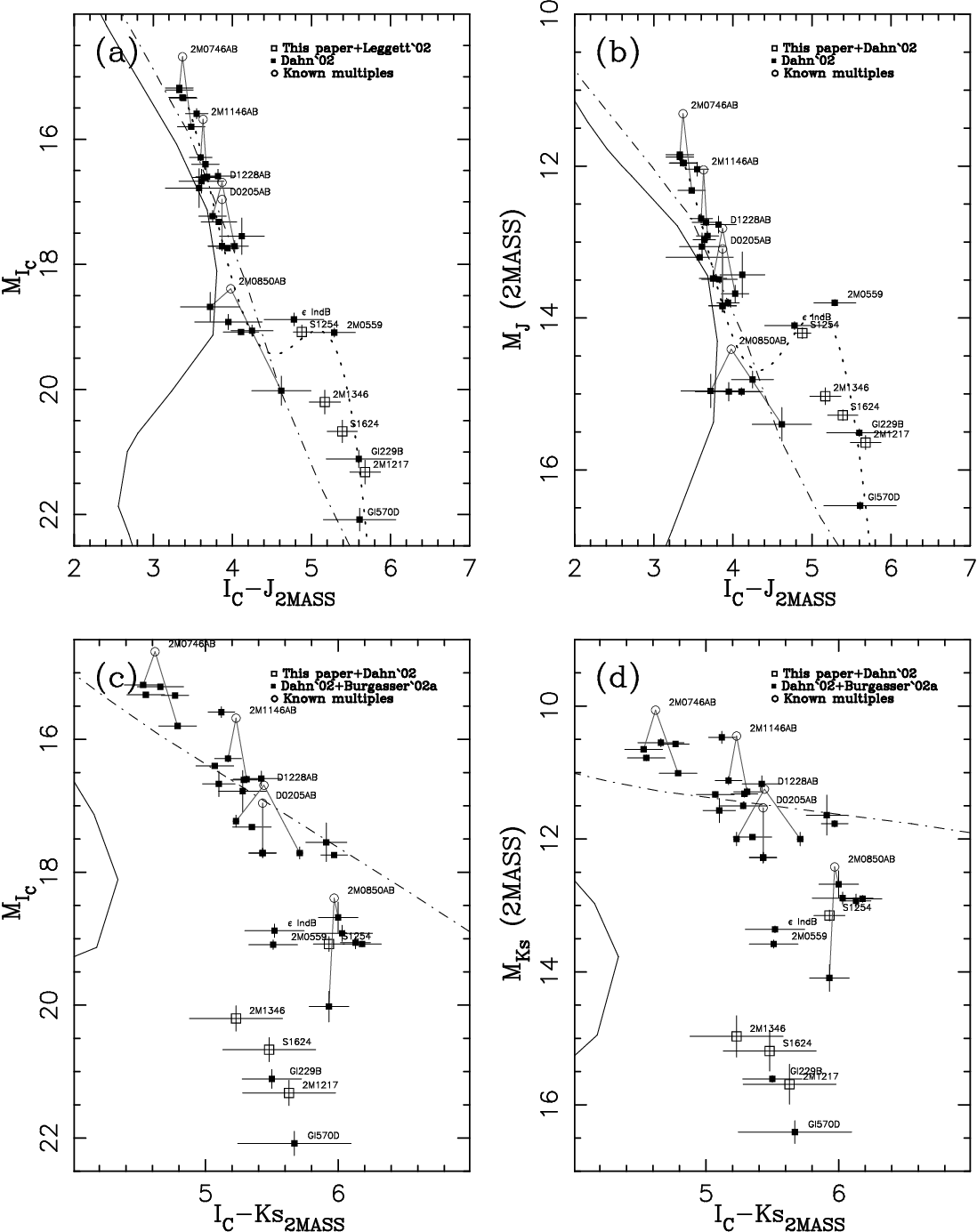}
   \caption{Colour magnitude diagrams based on  I$_C$ and 2MASS J,Ks photometry 
   \citep{da2002,burg_thesis}), and the parallaxes of this paper and \citet{da2002,scholz2003}. Symbols are as
   for Fig. \ref{abs_spt}.
   The solid lines are 1\,Gyr isochrones from the COND models of \citet{ba2003},
   and the dot-dashed lines are the 1\,Gyr isochrones from the DUSTY models of \citet{chab2000,all2001}. Both
   sets of models are have been kindly recalculated in the 2MASS \& I$_C$ filter systems by I.Baraffe.
   The dotted lines in panels a and b are high-order polynomial fits to the data -- see Section \ref{relations}.
    The decomposed J and K magnitudes of 2M0850B are much more uncertain than the formal
   uncertainties plotted would indicate, and should be treated as indicative only.}
   \label{cmd_iz}
\end{figure*}

\begin{figure*}
   \figurenum{\ref{cmd_iz}{\em (cont.)}}
   \centering\includegraphics[width=170mm]{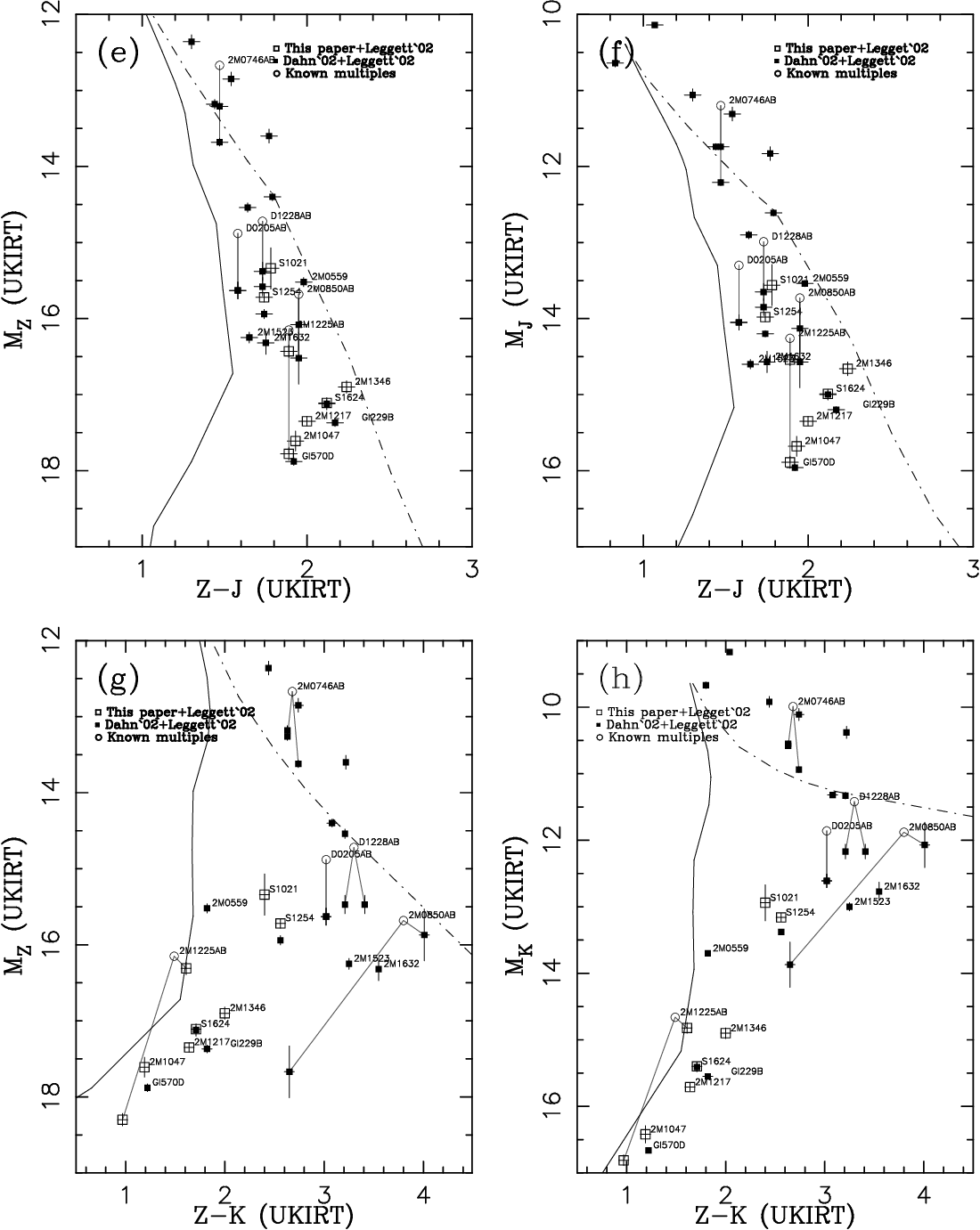}
   \caption{Colour magnitude diagrams based on UKIRT Z,J,K photometry \citep{le2002}),
   and the parallaxes of this paper and \citet{da2002,scholz2003}. Symbols are as
   for Fig. \ref{abs_spt}. The solid lines are 1\,Gyr 
   isochrones from the COND models of \citet{ba2003},
   and the dot-dashed lines are the 1\,Gyr isochrones from the DUSTY models of \citet{chab2000,all2001}. Both
   sets of models are have been kindly recalculated in the UKIRT filter system by I.Baraffe.
   The decomposed J and K magnitudes of 2M0850B are much more uncertain than the formal
   uncertainties plotted would indicate, and should be treated as indicative only.}
   \label{cmd_iz2}
\end{figure*}

\begin{figure*}
   \centering\includegraphics[width=170mm]{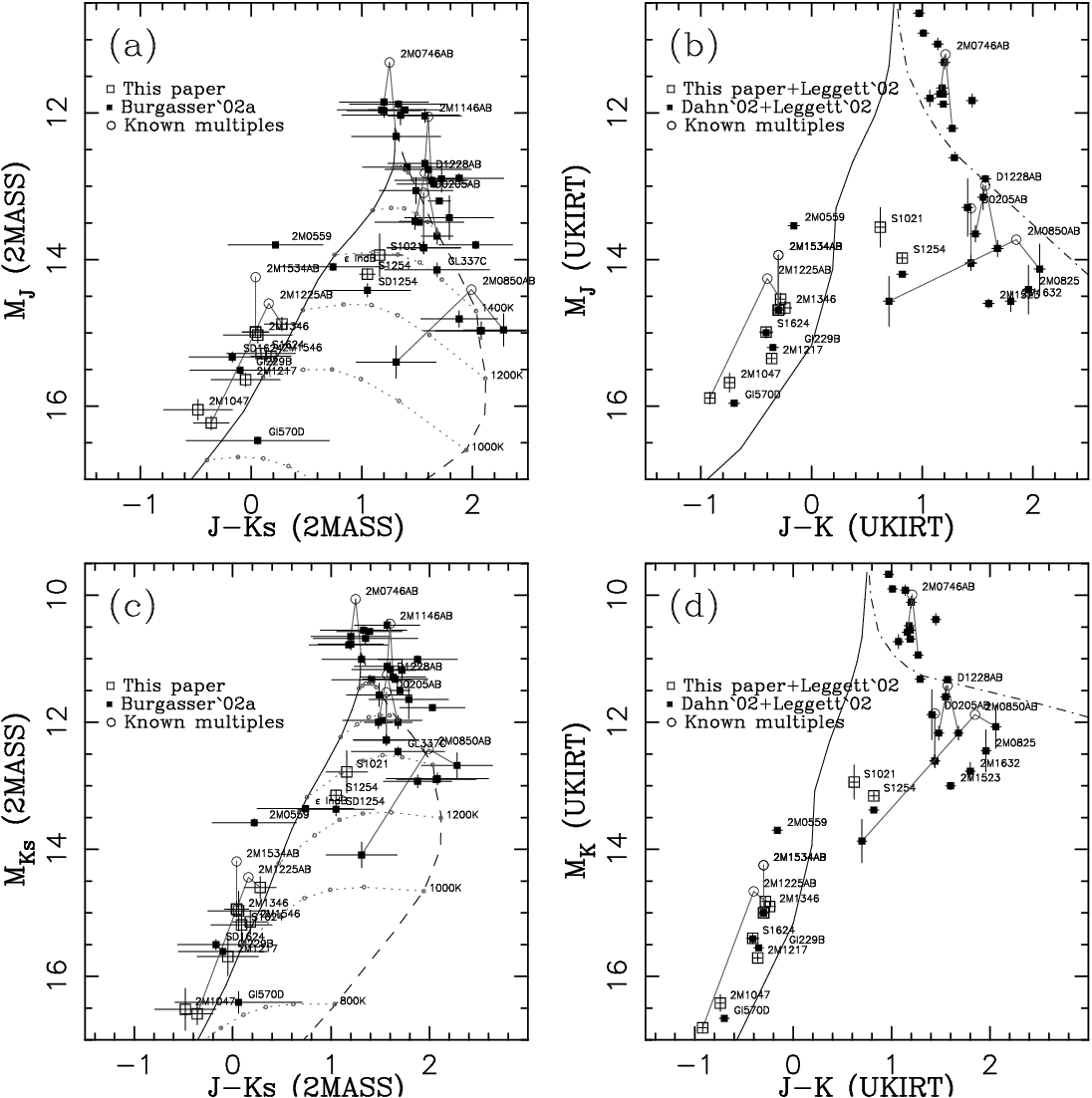}
   \caption{Colour magnitude diagrams using J-K colours on the 2MASS (panels a and c \citealt{da2002,burg_thesis})
   and UKIRT (panels b and d, \citealt{le2002} systems. Parallaxes from this paper and \citet{da2002,scholz2003}.
   The isochrones shown in panels (b) and (d) are
   1\,Gyr DUSTY  {\em (dot-dashed line)} and COND {\em (solid line)} models of
   \citet{ba2003,chab2000,all2001}.
   The isochrones in panels (a) and (c) are the CLEAR {\em (solid line)} and 
   CLOUDY {\em (dashed line)} models presented in \citet{bu2002b}. Joining these are dotted lines for
   the ``Cloud Clearing'' interpolations between these models with small circles highlighting the 20\%,
   40\%, 60\% and 80\% interpolation levels at effective temperatures of 800, 1000, 1200, 1400, 1600 \& 1800K.
   The decomposed J and K magnitudes of 2M0850B are much more uncertain than the formal
   uncertainties plotted would indicate, and should be treated as indicative only.}
\label{cmd_jk}
\end{figure*}


Using the same photometry, we can construct a variety of colour-magnitude diagrams. Figure \ref{cmd_iz}
shows such diagrams based around Cousins I$_c$, UKIRT Z and both UKIRT and 2MASS 
J,K photometry, while Figure \ref{cmd_jk} shows
similar diagrams for UKIRT and 2MASS J--K colours. The most noticeable feature of
these diagrams is how few are actually {\em useful} as traditional colour-magnitude
diagrams -- almost none show the simple monotonic relationships between absolute magnitude
and colour which hold for stars and brown dwarfs down to the early L dwarfs. 

Fig \ref{cmd_iz}c,d show that I--K colours jumps to the blue by I--K$\approx$0.5 mag
as the L-T transition is crossed  at M$_I$$\approx$19, M$_K$$\approx$ 13, but then 
tends redward again for later and later T dwarfs. However, Gl\,570D, one of the latest
and faintest T dwarfs currently known, never becomes as red as the latest L dwarfs. 
This blueward jump  is particularly pronounced at M$_I$ where the absolute magnitudes of
L8 and early T dwarfs are indistinguishable. As a result I--K should be considered
a poor indicator for determining the absolute magnitude or effective temperature of 
late L  to late T dwarfs. In particular, and luminosity function based on I--K$\ga$5
will be subject to serious biases which will introduce completely spurious structure
into the luminosity function.

Fig \ref{cmd_iz}a,b shows that I--J colour-magnitude diagrams can be
considered the ``best of a bad bunch'' when it comes to the traditional use of
colour-magnitude diagrams (i.e. estimating absolute magnitudes from photometric colours),
since the cooling curves of brown dwarfs do not reverse in I--J as they
do for every other panel of Figs \ref{cmd_iz} and \ref{cmd_jk}. Even so,
between I--J=4 and I--J=5 they show the same pronounced ``S-curve''  seen in the
spectral type data of Fig. \ref{abs_spt}, with early T dwarfs being up to a magnitude
brighter in M$_J$ than late L dwarfs. And, once again we see that 2M0559 appears
anomalously bright, suggestion binarity in spite of \citet{bu2003a}'s failure
to resolve it with HST..

The Z--J colour-magnitude diagrams (Fig.\ref{cmd_iz}e,f) reveal a very steep colour-magnitude relation,
with scatter which is significantly larger than the photometric errors. The slope
of the colour-magnitude relation is so steep that no meaningful estimate of M$_Z$ or M$_J$ 
can be derived from a Z--J colour. This is not surprising, given the very close
effective wavelengths of HgCdTe-based Z and J photometry.
There is some evidence for trend at the bottom
of this colour-magnitude diagram that at M$_Z \ga 16.5$ and M$_{Ks} \ga 14.5$,  Z--J
colours becomes {\em bluer} for fainter and later-type objects. 

Colour magnitude diagrams involving Z--K (Figs \ref{cmd_iz}g-h) and J--K (Figs \ref{cmd_jk}a-d)
show an especially pronounced reversal of the brown dwarf cooling curves beyond M$_K \approx 12.5$, M$_J \approx 14$ and
M$_Z \approx 16$. This has been noted in J--K colour-magnitude diagrams by several
authors (e.g. \citealt{bu2002b}). Major changes take place in photospheres below the L-T transition, 
with the result that T dwarfs
swap from very red, to very blue, J--K colours. It is interesting that the Z--K diagrams show
almost identical behavior, with Z--K colours for the very faintest T dwarfs becoming 
as blue as Z-K$\approx$-1.
This compares with colours based on the SDSS $z^\prime$ filter (see eg.\citealt{da2002} Fig. 3) which
continue to become redder for the latest T dwarfs. This once again clearly demonstrates the considerable
difference between    CCD-based $z^\prime$, and the HgCdTe-based Z.

There is a clear warning to astronomers implicit in these diagrams -- conclusions reached about
luminosity- and mass-functions based on luminosity and/or colours for the L-T effective
temperature range are fraught with difficulty. In  particular, luminosity functions determined
from the colours of objects in field samples will produce completely spurious features in
the derived luminosity- and mass-functions, unless the various ``bumps and wiggles''
in these diagrams are adequately and correctly modeled.
(See for example \citet{rg1997}'s demonstration of the formation of a ``false'' peak in M dwarf
luminosity function based on the traditional -- and inadequate -- parametrization
of the M dwarf colour-magnitude relation). 
Similarly, determining bolometric luminosity functions from apparent magnitudes
in cluster-based samples is problematic, as we can expect similar ``bumps
and wiggles" to be present in the bolometric correction relations for the L  and T dwarfs. 

Features like these can introduce significant {\em systematic} biases into
the mass functions derived from even a perfect statistical sample. 
{\em Actual} statistical data with all the added
complexities of uncertain age and binarity distributions add yet more
complications. Monte Carlo simulations are
essential to the interpretation of any luminosity- or mass-function in the L-T
effective temperature range. it is important to carefully
``reverse'' model such functions from sets of mass-function models, through a variety of possible
colour-magnitude and bolometric-correction relations (as allowed by the extant data), 
to sample observational data. Then such artificial data can then be  meaningfully compared
to statistical samples {\em in the observational plane}.
Mass- or luminosity functions  which do {\em not} include such extensive
reverse modeling  should be treated with the utmost suspicion.

\subsection{Theoretical Models for L  and T dwarfs}

Ultra-cool dwarfs are notoriously difficult to model -- the components which
need to be included in models for L  and T dwarfs include \citep{all1997}: 
the effects of tens of millions
of molecular transitions in species including H$_2$O, CH$_4$, TiO, VO, CrH, FeH, and a host of others;
complex treatments of the line wings of enormously
H$_2$ and He pressure-broadened neutral alkali lines like K and Na, Rb and Cs;
collision-induced molecular H$_2$ opacity;
both the chemistry and opacity involved in the condensation, settling, revapourisation 
and diffusion of a variety of condensates
including liquid Fe, solid VO, and a range of aluminium, calcium, magnesium and
titanium bearing refractories;  and
finally (and least readily modeled of all) the effects of rotation-induced weather
on the cloud decks which condensates will form.

Significant progress has been made in recent years on the detailed solution of
photospheric models using very large line lists (see \citet{all1997} for a review). 
Probably the largest
outstanding problem for modelers of L  and T dwarfs is dealing with condensation.
Three approximations to this complex situation have currently been implemented.
``Dusty'' models (eg. the DUSTY model of \citealt{all2001})
assume condensates remain well suspended and in chemical equilibrium where they form
in the photosphere. In general such models have been shown to work reasonably well for
L dwarfs, suggesting that their cloud layers lie within their photospheres.
``Condensation'' models (eg. the COND models of \citealt{ba2003}, and the CLEAR models
of \citealt{bu1997}) neglect dust opacities, to simulate the removal of all condensates
from the photosphere as they form (most likely through gravitational settling).
The ``CLOUDY'' models of \citet{am2001} and \citet{marley2002} incorporate
a model for condensate cloud formation, based on an assumed sedimentation
efficiency parameter $f_{rain}$.

\subsubsection{Colour-magnitude diagrams in J--K and Z--K}

Figures \ref{cmd_iz} and \ref{cmd_jk} have over-plotted on them a variety of these models,
including the DUSTY \citep{chab2000,all2001} and COND models \citep{ba2003} for an age
of 1\,Gyr,
and the CLEAR and CLOUDY models as presented in \citet{bu2002b} for $f_{rain}=3$ (determined
as the best fit for this model in Jupiter's ammonia cloud deck \citealt{marley2002}). As previous
studies have shown, DUSTY models reproduce the general features
(if not the precise colours) of the cooling curves for L dwarfs, but then
proceed to much redder colours than are observed beyond L8. This 
has been interpreted as indicating that condensates are present in the photosphere
of L dwarfs. The
COND and CLEAR models reproduce the general features of the cooling curves for
late T dwarfs, indicating that at these effective temperatures condensate opacities
do not contribute to the radiative transfer, which suggests that the condensate layers
have dropped below the photosphere. DUSTY and CLEAR/COND models, therefore, describe
the  ``boundary conditions'' to the condensate opacity problem, and are appropriate for the
L  and late T types respectively. But what about the intermediate case which must be
appropriate to early T dwarfs? This is exactly the situation with the sedimentation
models of \citet{marley2002} should be able to address.

\citet{bu2002b} compared their CLOUDY models for $f_{rain}=3$ with a 2MASS M$_J$:J-K 
colour-magnitude diagram (as we do in Fig. \ref{cmd_jk}). As for the DUSTY models,
the CLOUDY models predict the general behaviour of L dwarfs, and then veer towards
bluer J--K colours at late T dwarf temperatures. However, this transition does not match
the observed sequence, which transitions nearly horizontally
between the L dwarf/DUSTY/CLOUDY sequence and late-T dwarf/CLEAR/COND model 
 at M$_J$$\approx$14, M$_{Ks}$$\approx$13. (We note that though the equivalent models are not available
 in the UKIRT Z,J,K bandpasses, very similar behaviour is seen in Fig. \ref{cmd_iz}, with a clear
 transition between the L dwarf and late T dwarf sequences.)
\citet{bu2002b} suggested that a possible resolution for this discrepancy could be 
the appearance of uneven cloud cover on the surface of early T dwarfs.
This would allow the emergent spectrum to appear as a ``mixture'' of the CLEAR/COND and
CLOUDY spectra. They modeled this by interpolating between their CLEAR and CLOUDY models
at effective temperatures of 800\,K, 1000\,K, 1200\,K, 1400\,K, 1600 and 1800\,K with varying
fractions of the two models (ie. 20\%,40\%,60 \& 80\%). The tracks for these
``mixture'' models are shown in Fig. \ref{cmd_jk} as dotted lines, and suggest that
there is a transition sequence between L  and T dwarfs at T$_{\mathrm eff}$ $\approx$ 1300K.
SDSS1021, SDSS1254, 2M1225, $\epsilon$\,Ind\,B and possibly 2M0850B (though with some uncertainty
because of the poor quality of its decomposed secondary flux) fill out this
transition region. The status of 2M0559 is unclear. If it is a single object, then it probably
represents the `top' of the late T dwarf cooling sequence, which is $>$1 mag. brighter in
M$_J$ than the bottom of the L dwarf sequence. If however, it is a binary, then the 
prototype for the `top'
of the late T dwarf cooling sequence is probably more like the object 2M1225A or 2M1346
at a spectral type of T5.5-T6. The ``transition temperature'' indicated by the additional T dwarfs in
this work is slightly warmer ($\approx$1300K) than that found by \citet{bu2002b}.

An alternative dust model to the sedimentation
models of Marley et al. has been developed by \citet{tsuji2003}. These
``Unified Cloudy Models'' (UCM) are built around a single thin dust layer
in which particles of size greater than a critical radius are removed from
the photosphere by sedimentation. This critical radius is parametrized by
a critical temperature $T_{cr}$ below which dust particles sediment, which
is determined by comparing model results to colour-magnitude diagrams. This
single model has the advantage of predicting the gross behaviour of brown dwarfs
as they transition from L  to T spectral types, with a single model.
Unfortunately, \citep[Fig.2]{tsuji2003} the detailed behaviour of the models
does not match observations. In particular, UCM cannot make L dwarf as red
or faint in M$_J$:J--K as they actually appear. Nor does it predict the observed
brightening of the ``early T dwarf hump'' other than as an age-selection
effect, which we conclude below, is not the case.

It should be noted, however, that the interpretation of Marley et al. models
and data in Fig. \ref{cmd_jk} in terms of 
cloud openings (i.e. as providing evidence for the existence of weather in
early T dwarfs) is quite dependent on the details present in the
\citet{marley2002} models. An independent test of this conclusion is clearly desirable.
Fortunately, Fig. \ref{cmd_jk} indicates that J and K band time-series photometry can
provide that test. The location of a given object on the ``transition sequence'' will
depend critically on its fractional cloud cover. Because this could be expected to change
as each brown dwarf rotates, a statistical study of the
J-band variability from late L dwarfs to late T dwarfs should find
stronger variability in early T dwarfs, than in late L dwarfs or late T dwarfs.

Finally, we note that although the COND models do not do a very good job of
predicting the {\em absolute} colours of late T dwarfs in Z--J and Z--K (Fig. \ref{cmd_iz}e-h),
they do suggest a trend for late T dwarfs to become bluer in both Z--J and Z--K
as they get colder and fainter than M$_J$$\approx$14.5 and M$_K$$\approx$14.75.
Moreover, the available data suggest this trend is real, though the absolute
colours of T dwarfs at these magnitudes are somewhat redder than the models would predict.

\subsubsection{The ``Early T Hump''}
\label{sec_hump}
\begin{figure*}
   \centering\includegraphics[width=170mm]{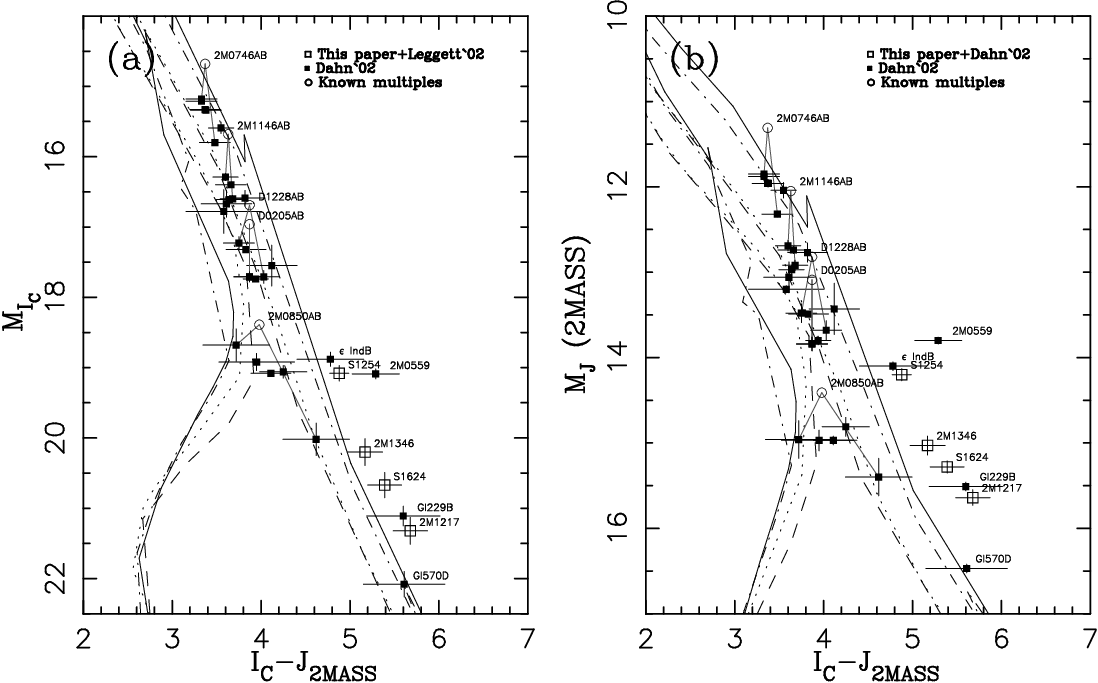}
   \caption{Colour magnitude diagrams based on  I$_C$ and 2MASS J photometry 
   \citep{da2002,burg_thesis}), and the parallaxes of this paper and \citet{da2002,scholz2003}.
   DUSTY \citep{chab2000,all2001} (rightmost lines) and COND \citet{ba2003} (leftmost lines)
   are shown as in Fig. \ref{cmd_iz}, but now for a range of ages: 50\,Myr {\em (solid)}, 
   100\,Myr {\em (dot-dashed)}, 1\,Gyr {\em (dotted)}, 5\,Gyr {\em (dashed)}.}
   \label{cmd_age}
\end{figure*}


The M$_J$:I$_C$--J colour-magnitude diagrams shown in Fig. \ref{cmd_iz}a-b indicate a remarkable
brightening at M$_J$ for the observed early T dwarfs. Unfortunately, neither the COND nor the
DUSTY models indicate why this should be so. The DUSTY models predict an extension of the
L dwarf sequence, which we have good reason to believe is not correct, based on the
analysis of colour-magnitude diagrams in J--K above. Unfortunately, the COND models {\em also}
fail to look even remotely like the available data for T dwarfs in Fig. \ref{cmd_iz}a-d. Shortcomings in
these models at short wavelengths have been noted by \citet{ba2003}, which are thought to
be due to an inadequate treatment of the extremely broad wings of the K and Na lines
at these wavelengths.

One possible interpretation of the ``early T hump'' in Fig. \ref{cmd_iz}a-b, is that
it could be a gravity effect \citep{tsuji2003}. Very young brown dwarfs will have isochrones slightly
offset to brighter magnitudes than older brown dwarfs, because of their lower gravities.
This effect is particularly pronounced in photospheres in which dust is an important opacity source.
It is possible then that the ``early T hump'' could be produced by the preferential
selection of young, bright brown dwarfs.

Fig. \ref{cmd_age} plots the same data as that shown in Fig. \ref{cmd_iz}a-b, but now
we plot four isochrones spanning 50\,Myr-5\,Gyr to examine the effects of age. 
The figure shows that, as expected, the DUSTY models (most appropriate for
L dwarfs) show significant offsets in their isochrones of a magnitude or more between
50\,Myr and 5\,Gyr. These offsets are not as marked for the
COND models. Unfortunately, interpreting the ``early T hump'' as an age effect is 
severely complicated by the fact that it occurs at {\em exactly} the point where there is good
evidence to believe neither the DUSTY or COND models are working.

For the L dwarfs and late T dwarfs, the spread in the colour-magnitude diagram
is not pronounced, (particularly when known binaries are decomposed), 
suggestive of the small 100\,Myr -- 1\,Gyr age spread seen in other studies of L 
and T dwarfs \citep{da2002,scholz2003}. It is certainly nowhere near as pronounced as
would be necessary to produce the age spread required to account for the more than
1\,magnitude brightening of the``early T hump'' all by itself. 
Moreover, there is a definite spectral type trend {\em along} the
track represented by the ``early T hump'', as seen in the spectral-type-magnitude diagrams of
Fig. \ref{abs_spt} -- from the late L dwarfs, through $\epsilon$\,Ind\,B,
SD1254, SD1021 to 2M0559. This same trend is seen in the colour-magnitude diagrams.
We interpret this as indicating that the ``early T hump'' truly is a feature
in the cooling curve of brown dwarfs, rather than an artifact of youth and selection.

%
%
\subsection{The Onset of CH$_4$ absorption in Clusters}
\label{ch4}

\begin{figure*}
   \centering\includegraphics[width=170mm]{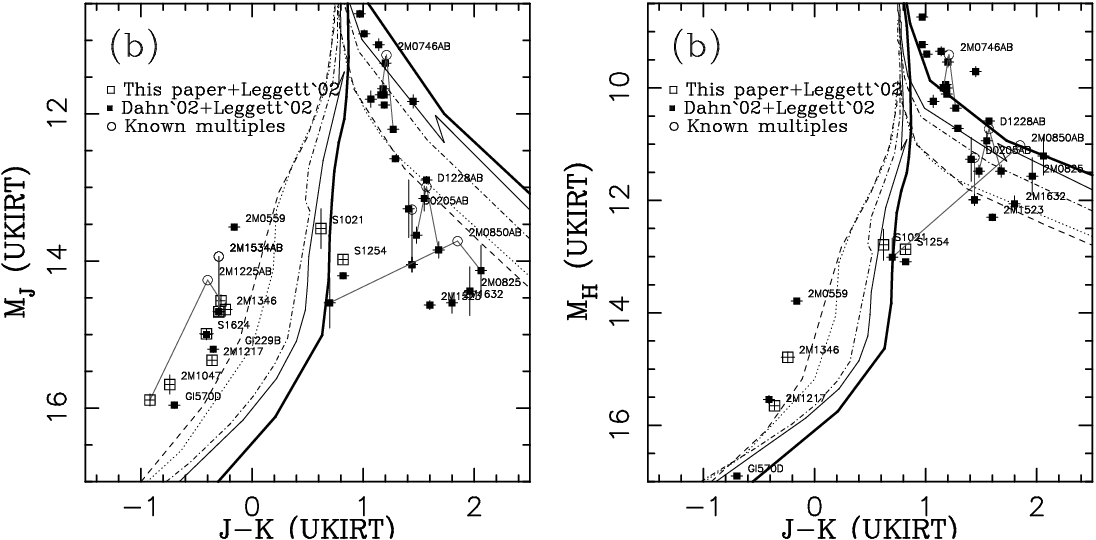}
   \caption{Colour magnitude diagrams in UKIRT M$_J$ and M$_H$ vs J--K.
   DUSTY \citep{chab2000,all2001} (rightmost lines) and COND \citet{ba2003} (leftmost lines)
   are shown as in Fig. \ref{cmd_iz}, but now for a range of ages: 10\,Myr,(heavy solid) 50\,Myr {\em (solid)}, 
   100\,Myr {\em (dot-dashed)}, 1\,Gyr {\em (dotted)}, 5\,Gyr {\em (dashed)}.
   The decomposed J and H magnitudes of 2M0850B are much more uncertain than the formal
   uncertainties plotted would indicate, and should be treated as indicative only.}
   \label{cmd_h}
\end{figure*}

Methane filters centered on the strong CH$_4$ absorption bands in the H-band
have been acquired by a number of observatories for use in their infrared
cameras. Given we have now measured just where, in absolute magnitude,
the T dwarf class occurs, the question arises, ``At what magnitudes will
CH$_4$ absorption in young star clusters set in?'' Fig. \ref{cmd_h} shows
UKIRT M$_J$ and M$_H$:J--K colour-magnitude diagrams, along with the DUSTY and COND models
at ages from 10\,Myr to 5\,Gyr. Based on these diagrams,
we can conclude that for field T dwarfs, as discovered by the 2MASS and SDSS
surveys, CH$4$ absorption (corresponding to spectral classes around T2 and later)
sets in at at M$_H$$\approx$13, and somewhat more confusingly at
M$_J$$\sim$14  -- though because of the brightening of ``early T hump''
at J non-CH$_4$-absorbing L dwarfs will actually be fainter than the earliest
CH$_4$ absorbing T dwarfs.

Because the turn on of CH$_4$ absorption is primarily an effect driven
by effective temperature, to {\em first order} it will occur at the
same absolute magnitude in young clusters as it does in the field.
Looked at in slightly more detail, however, we can see that for a given
colour in Fig. \ref{cmd_h}, there is a small offset to brighter magnitudes
 for younger objects -- in the DUSTY atmosphere case a 10\,Myr dwarf
at the end of the L8 sequence will be $\approx$1.0\,mag. brighter than a 1\,Gyr dwarf
of the same colour and effective temperature, and 1.3\,mag brighter than a 5\,Gyr dwarf.
In the COND case the equivalent differences are 1.3 and 1.7 magnitudes.
The likely ages for our field T dwarfs will be somewhere in the range 100\,Myr-1\,Gyr
\citep{da2002,scholz2003}. This would suggest that in clusters like IC\,2391 or IC\,2602
of age 10-20\,Myr at d$\approx$150\,pc the absolute magnitude for CH$_4$ onset 
will be M$_H$$\sim$12-12.5,  or equivalently H$\sim$18-18.5. For older
clusters like the Pleiades (100\,Myr, d$\sim$125\,pc) these numbers are more
 like M$_H$$\sim$12.5-13 or H$\sim$18-18.5.Both of these are eminently reachable magnitude limits
 with wide-field cameras on 4m-class telescopes, suggesting that CH$_4$ imaging may
 be a powerful tool for easily conducting an unbiased census of T dwarfs in 
 large open clusters.
  Similarly for more compact, but distant clusters, like Trapezium ($\sim$25\,Myr, 450\,pc)
 observations at H$\sim$20-20.5 are tractable over the fields-of-view required on 8m-class
 telescopes.

Fig. \ref{cmd_h} also has implications for the interpretation of potential cluster
membership. For example, \citet{zo2002} have found a T6 dwarf in the direction
of the $\sigma$\,Orionis cluster. Fig. \ref{cmd_h} suggests that a field brown dwarf
of this spectral type will have M$_H$=15.0$\pm$0.5. For the much younger age of
$\sigma$\,Orionis (1-8\,Myr \citealt{zo2002}) this will be more like M$_H$=14.0$\pm$0.5,
which would imply a distance to the $\sigma$\,Ori\,J053810.1-023626 T6 dwarf of 192$\pm$50\,pc --
more consistent with being a foreground object than a member of the cluster at d=352\,pc
\citep{perry1997}.

\subsection{Colour-Magnitude \& Spectral-Type-Magnitude relations for L  and T dwarfs}
\label{relations}

Figures \ref{abs_spt} and \ref{cmd_iz} have over-plotted on some of
their panels high-order polynomial fits to the weighted (and
binary decomposed) data. As inspection of the figures shows, these fits are not always
particularly successful at modeling the extremely complex behaviour of these cooling curves
in the observed passbands. Nonetheless, in the absence of working atmospheric models
the fits may be a useful tool, so long as their weaknesses
are acknowledged. We therefore provide the coefficients for these fits,
and the root-mean-square scatter about the fits in  Table \ref{fits}.

\begin{table*}
 \centering
 \begin{minipage}{190mm}
 \scriptsize
 \caption{Polynomial coefficients for plotted fits \label{fits}}
  \begin{tabular}{llcrrrrrrrr}
  \tableline\tableline
   $P(x)$ &  $x$\footnote{SpT $= i$ for M$i$, $=j+10$ for L$j$ and $=z+19$ for T$z$ spectral types on the \citet{ki1999} 
   system for M  and L dwarfs, and the \citet{bu2002a} system for T dwarfs.}  
   & RMS & $c_0$\footnote{$P(x) = c_0 + c_1\,x + c_2\,x^2 ...$}   
   & $c_1$  & $c_2$ & $c_3$ & $c_4$ & $c_5$ & $c_6$ 
   & $c_7$ \\
   \tableline
   M$_{Ks}$ (2M)&SpT   &0.38 & 6.27861e$+$1 & -1.47407e$+$1 & 1.54509e$+$0 & -7.42418e$-$2 & 1.63124e$-$3 & -1.25074e$-$5 &   -   & -     \\
   M$_K$ (U)&SpT       &0.40 & 8.14626e$-$1 & 2.95440e$+$0 & -3.89554e$-$1 & 2.68071e$-$2 & -8.86885e$-$4 & 1.14139e$-$5  &   -   & -     \\
   M$_J$ (2M)&SpT      &0.36 & 8.94012e$+$2 & -3.98105e$+$2 & 7.57222e$+$1 & -7.86911e$+$0 & 4.82120e$-$1 & -1.73848e$-$2 & 3.41146e$-$4 & -2.80824e$-$6\\
   M$_J$ (U)&SpT       &0.30 & 5.04642e$+$1 & -3.13411e$+$1 & 9.06701e$+$0 & -1.30526e$+$0 & 1.03572e$-$1 & -4.58399e$-$3 & 1.05811e$-$4 & -9.91110e$-$7\\
   M$_{Ic}$&SpT        &0.37 & 7.22089e$+$1 & -1.58296e$+$1 & 1.56038e$+$0 & -6.49719e$-$2 & 1.04398e$-$3 & -2.49821e$-$6 &   -   & -     \\
   M$_Z$ (U)&SpT       &0.29 & 4.99447e$+$1 & -3.08010e$+$1 & 8.96822e$+$0 & -1.29357e$+$0 & 1.02898e$-$1 & -4.57019e$-$3 & 1.05950e$-$4 & -9.97226e$-$7\\
   M$_{Ic}$&I$_C$--J   &0.67 & 1.17458e$+$3 & -1.22891e$+$3 & 5.00292e$+$2 & -9.70242e$+$1 & 8.86072e$+$0 & -2.97002e$-$1 &   -   & -     \\
   M$_J$ (2M)&I$_C$--J &0.63 & 1.52199e$+$3 & -1.69336e$+$3 & 7.41385e$+$2 & -1.58261e$+$2 & 1.64657e$+$1 & -6.66978e$-$1 &   -   & -     \\
   \tableline                                    
\end{tabular}                             
\end{minipage}                            
\end{table*}                              


\section{Conclusion}

We have shown that high precision parallaxes can be obtained with
common-user near-infrared cameras using techniques very similar to those
used in optical CCD astrometry. The new
generation of large format infrared imagers based on
HAWAII1 (1K) and HAWAII2 (2K) HgCdTe arrays,
and the new generation of large format InSb arrays, offer exciting prospects
for the astrometry of cool brown dwarfs in the future.
Due to their infrared methane absorption bands, T dwarfs have quite similar effective wavelengths
to the ensemble of background reference stars, which makes the 
correction of differential colour refraction effects considerably easier.
The ``early T hump'' (i.e. the brightening at the J band of early T dwarfs relative
to late L dwarfs) appears to be a feature of brown dwarf cooling curves,
rather than an effect of binarity or age. And finally, these data imply
that detection of T dwarfs in clusters can be made directly at tractable magnitudes
in the H-band, opening the way to a new generation of cluster mass function studies
based on the powerful technique of CH$_4$ differential imaging.

\acknowledgements

The authors gratefully acknowledge support for this program from the
Australian Goverment's Access to Major Research Facilities Program (grants 99/00-O-15 \& 01/02-O-02),
and the AAO Director Dr B.Boyle. 
CGT would like to thank Dr Joss Hawthorn for his assistance with
an early draft. 
Publication funds were
provided through support for HST proposal 8563, and AJB acknowledges 
support by the National Aeronautics and Space
Administration (NASA) through Hubble Fellowship grant HST-HF-01137.01. 
Both are provided by NASA through a
grant from the Space Telescope Science Institute, operated by the Association
of Universities for Research in Astronomy, Inc., under NASA contract
NAS5-26555. JDK acknowledges the 
support of the Jet Propulsion Laboratory, California Institute of Technology,
which is operated under contract with NASA.
This publication makes use of data from the Two Micron All Sky
Survey, which is a joint project of the University of Massachusetts and the
Infrared Processing and Analysis Center, funded by NASA and the NSF.
Finally, we would like to thank Dr Hugh Harris for a helpful and efficient
referee's report.

\label{lastpage}

\end{document}